\documentclass[10pt,journal,compsoc]{IEEEtran}
\ifCLASSOPTIONcompsoc
  \usepackage[nocompress]{cite}
\else
  \usepackage{cite}
\fi

\ifCLASSINFOpdf
  \usepackage[pdftex]{graphicx}
  \graphicspath{{figures/}}
\else
\fi
\usepackage{amsmath}
\usepackage{url}

\usepackage[hidelinks]{hyperref}
\usepackage{booktabs} 
\usepackage{tabularx}
\usepackage{multirow}
\usepackage[table, dvipsnames]{xcolor}
\usepackage{footnote}
\usepackage{xspace}
\usepackage{color,soul} 
\usepackage{enumitem}
\usepackage{comment}
\usepackage{tabu}
\makesavenoteenv{tabular}

\usepackage{tikz}
\newcommand*\circled[1]{\tikz[baseline=(char.base)]{\node[shape=circle,draw,inner sep=0.5pt] (char) {#1};}}

\def\markup{0}
\if\markup1
\newcommand{\rv}[1]{{\leavevmode\color{RoyalBlue}#1}}
\else
\newcommand{\rv}[1]{#1}
\fi

\sethlcolor{Goldenrod}

\newcommand{\name}{ChartStory\xspace}
\newcommand{\baseline}{Baseline\xspace}

\newcommand{\q}[1]{\textit{``#1''}}
\newcommand{\p}[1]{\textit{P#1}}
\newcommand{\e}[1]{\textit{E#1}}

\newcommand{\etal}{et~al.}
\newcommand{\eg}{e.g.}
\newcommand{\ie}{i.e.}

\begin{document}
%
\title{\name{}: Automated Partitioning, Layout, and Captioning of Charts into Comic-Style Narratives}
%
%
%
%

\author{Jian Zhao$^{\dagger,\star}$, Shenyu Xu$^\dagger$, Senthil Chandrasegaran$^\dagger$, Chris Bryan, Fan Du, \\Aditi Mishra, Xin Qian, Yiran Li, and Kwan-Liu Ma
\IEEEcompsocitemizethanks{
\IEEEcompsocthanksitem $\dagger$ These authors contributed equally; $\star$ Corresponding author.
\IEEEcompsocthanksitem J. Zhao is with the University of Waterloo. E-mail: jianzhao@uwaterloo.ca.
\IEEEcompsocthanksitem S. Xu is with the Georgia Institute of Technology. E-mail: shenyuxu@gatech.edu.
\IEEEcompsocthanksitem S. Chandrasegaran is with the Technische Universiteit Delft. E-mail: r.s.k.chandrasegaran@tudelft.nl.
\IEEEcompsocthanksitem Y. Li and K.-L. Ma are with the University of California, Davis. E-mail: \{ranli, klma\}@ucdavis.edu
\IEEEcompsocthanksitem C. Bryan and A. Mishra are with the Arizona State University: \{chris.bryan, amishr45\}@asu.edu.
\IEEEcompsocthanksitem F. Du is with Adobe Research. E-mail: fdu@adobe.com.
\IEEEcompsocthanksitem X. Qian is with the University of Maryland. E-mail: xinq@umd.edu.
}
\thanks{Manuscript received XXXX; revised XXXX.}
}

%
%

\markboth{IEEE Transactions on Visualization and Comptuer Graphics}%
{Zhao \MakeLowercase{\textit{et al.}}: \name}
%



\IEEEtitleabstractindextext{%
\begin{abstract}
  Visual data storytelling is gaining importance as a means of presenting data-driven information or analysis results, especially to the general public.
  This has resulted in design principles being proposed for data-driven storytelling, and new authoring tools being created to aid such storytelling.
  However, data analysts typically lack sufficient background in design and storytelling to make effective use of these principles and authoring tools.
  To assist this process, we present \textit{\name} for crafting data stories from a collection of user-created charts, using a style akin to comic panels to imply the underlying sequence and logic of data-driven narratives.
  Our approach is to operationalize established design principles into an advanced pipeline which characterizes charts by their properties and similarity, and recommends ways to partition, layout, and caption story pieces to serve a narrative.
  \name\ also augments this pipeline with intuitive user interactions for visual refinement of generated data comics.
  We extensively and holistically evaluate \name{} via a trio of studies. We first assess how the tool supports data comic \textit{creation} in comparison to a manual baseline tool. 
  Data comics from this study are subsequently compared and evaluated to \name's automated recommendations by a team of narrative visualization practitioners.
  This is followed by a pair of interview studies with data scientists using their own datasets and charts who provide an additional assessment of the system.
  We find that \name\ provides cogent recommendations for narrative generation, resulting in data comics that compare favorably to manually-created ones.
\end{abstract}

}

\maketitle

\IEEEdisplaynontitleabstractindextext

%
\IEEEpeerreviewmaketitle

\IEEEraisesectionheading{\section{Introduction}\label{sec:introduction}}

%
%
%
%


\begin{figure*}[!tb]
  \centering
  \includegraphics[width=\linewidth]{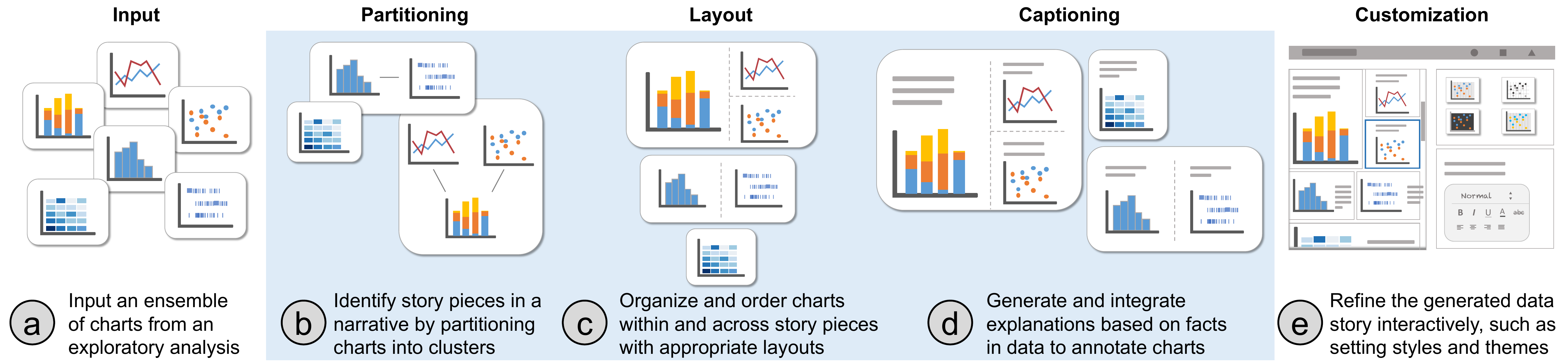}
  \vspace{-8mm}
  \caption{\name\ takes an ensemble of user-created charts (a) and automatically generates a data comic (b-d). 
    This is achieved through a back-end pipeline with operations to identify story pieces (b), organize \& order story pieces (c), and generate \& integrate explanations (d).
    The user can also interactively edit the captions and layout of the charts, and change the style or appearance of the data comic (e).
  }
  \label{fig:overview}
\end{figure*}

\IEEEPARstart{V}{}isual data storytelling concerns the communication of data insights and narratives to general audiences using engaging visualizations. 
\rv{The notion of ``visual data story''~\cite{lee2015more} includes the discovery of interesting information in data as ``story pieces,'' the representation of them using visualizations and annotations, and the sequencing of these representations into a narrative to communicate a high-level goal.
When designed well, visual stories can greatly improve the comprehension of data even among laypeople.}

\rv{However, the creation of visual data stories can be especially difficult for a general data analyst, who is \textit{familiar with the data}, but often has \textit{little background in art or design}.
Research on computational notebooks highlight the tension between \textit{exploration}---conducted at a personal scale with messy results---and \textit{explanation}---where results are cleaned up at the expense of provenance~\cite{Kery2017exploring, Rule2018exploration, Subramanian2020tractus}.
There is a need for an intermediate step to mitigate this tension.
One main challenge is identifying a narrative---converting the analysis results (\eg, a collection of charts) into a compelling story that resonates with the reader and reveals a logical progression of ideas.
The second is presenting the narrative---arranging the charts and textual annotations by leveraging design principles drawn from research on visual data storytelling.}

\rv{There exist theoretical frameworks to generate a narrative from a set of charts~\cite{lee2015more, hullman2013deeper}; yet, no satisfactory approach has been proposed to operationalize these theories. 
There also exist interactive authoring tools for presenting a narrative in the form of expressive visualizations~\cite{xia2018dataink, satyanarayan2014lyra}, annotations~\cite{ren2017chartaccent}, and comic-style storyboards~\cite{kim2019datatoon}; however, these tools still require much manual input and learning, not supporting automated narrative identification or presentation. 
In contrast, several works focus on automatically generating specific types of visual data stories, such as ``fact sheets,'' directly from data tables~\cite{Shi2020calliope,wang2019datashot,chen2018supporting}; however, they are mainly designed for general audience, not data analysts who wish to fully control the analysis and only automate the narrative generation.}   


\rv{To fill in the gap, we address the challenges of identifying and presenting a narrative through \name, a tool that helps analysts automatically generate \textit{data comics}---a major genre of visual data storytelling---from an ensemble of charts created during exploratory data analysis (\autoref{fig:overview}a).
As an initial step, we consider data comics, first defined by Zhao \etal\cite[p.~2]{zhao2015data}, as ``\emph{multiple visualizations (juxtaposed) into comic strip layouts consisting of a sequence of panels, each appropriately annotated and decorated with both visual and textual elements, and arranged into a sequence that progressively develops the overarching story told in the comic,}'' 
although traditional comic strips might include more embellishments.}
\name\ operationalizes a set of design requirements that we term \textit{Granularity}, \textit{Relatedness}, \textit{Explanation}, and \textit{Presentation} (GREP), synthesized from prior research in visual data storytelling~\cite{hullman2013deeper, lee2015more}. 
It generates data comics by identifying ``story pieces''~\cite{lee2015more} from the ensemble of charts (\autoref{fig:overview}b), organizing story pieces in a layout based on data comic design patterns~\cite{bach2018design} (\autoref{fig:overview}c), and generating natural-language captions to explain detected data facts (\autoref{fig:overview}d).

To achieve this automatic pipeline (\autoref{fig:overview}b-d), we leverage advanced analysis of chart specifications (such as marks, channels, and transformations) to algorithmically group and organize an ensemble of charts into appropriate layouts;
we also automatically extract data facts from the charts and use natural-language generation (NLG) techniques to annotate the data comics.
\name\ provides a set of easy-to-learn interactions that allows users to edit, customize, and refine of the generated data comic layout and captions (\autoref{fig:overview}e).
By operationalizing established principles for identifying and presenting narrative segments from charts, \name\ provides a good "first cut" of narration that the analyst can further edit and reorganize as needed.
While several aspects of \name\ are applicable to other storytelling mediums (e.g., infographics), we focus on data comics as they improve reader focus and engagement~\cite{wang2019comparing}.
We focus on the \textit{specific} type of data comics defined by Zhao \etal\cite{zhao2015data}.


We evaluate \name\ on its ability to aid comic-style narrative presentations of data analysis results from three aspects: \textit{authoring}, \textit{readability}, and \textit{overall use}.
The first, a controlled study, evaluates the ease of authoring, comparing \name\ against a baseline version with no automated grouping or layout abilities.
The study simulates the scenario of ``handoff'' in collaborative analysis~\cite{xu2018chart, Zhao2017supporting} where analysts share their analysis results with a team member who assimilates and presents the results.
The second study evaluates the readability of the grouping and layouts of data comics generated in the first study.
Finally, a third interview study with two data scientists show that---when using their own charts created from their own analyses---\name\ is able to convey a useful and presentable narrative.

In summary, our contributions in this paper include:
\begin{enumerate}[label=\textnormal{(\arabic*)}, nosep]
  \item A set of requirements (abbreviated as GREP) derived from prior work~\cite{lee2015more, hullman2013deeper, kim2017graphscape} to operationalize data-driven storytelling,
  \item An automated pipeline that operationalizes GREP, as well as an interactive system named \name{}, for partitioning, layout, and captioning an ensemble of charts to generate a data comic, and
  \item An evaluation that illustrates the value of our approach to both data comic creators and consumers.
\end{enumerate}

The code, associated data, and supplementary materials can be accessed at \url{https://github.com/WatVis/ChartStory}. 

\section{Background}


\subsection{Challenges in Visual Presentation of Data}
The importance of a seamless transition from analysis to presentation was emphasized almost right at the inception of visual analytics.
Thomas \& Cook recommended that effective tools require \emph{production}, \emph{presentation}, and \emph{dissemination}~\cite{thomas2005illuminating}.
Many of the existing tools focus mainly on the production, rather than the presentation and dissemination. 
For example, computational notebooks (\eg, Jupyter~\cite{kluyver2016jupyter}) are versatile media allowing the reproduction of analytic workflows.
However, sharing and presenting these analyses still requires exporting the results and displaying visualizations using a different format.

Some early research focuses on automated generation of charts to present relational information, exemplified by Mackinlay's APT~\cite{mackinlay1986automating}.
This focus still exists, as evidenced in recent and more sophisticated tools.
For instance, Voyager~\cite{wongsuphasawat2015voyager} provides a ranked recommendation of automatically-generated charts based on selected data variables.
CompassQL~\cite{wongsuphasawat2016towards} is a query language that groups similar visualizations and selects representatives from each group.
Draco~\cite{moritz2019formalizing} is a formal model that represents visualizations as facts and design guidelines as constraints. 
These tools focus on choosing the right visualization, encoding design knowledge and practices for creating visualizations.

Our main contributions lie in presentation and dissemination.
\name\ helps analysts communicate their findings as visual data-driven stories, using the comic-style medium.
We draw from theoretical and empirical studies in this area~\cite{lee2015more,hullman2013deeper,bach2018design}, and operationalize the extracted principles into an automatic pipeline that recommends partitioning, layout, and captioning when creating data comics.  

\subsection{Data-Driven Storytelling and Data Comics}

Analysis of a curated collection of recent stories~\cite{stolper2016emerging} presents a set of design intents: communicating narrative and explaining data, linking separated story elements, enhancing structure and navigation, and providing controlled exploration.
Several works focused on theoretical conceptualizations of these intents.
For instance, Generalized Space-Time Cubes~\cite{bach2017descriptive} displays both existing and unrealized temporal narrative designs as a projection, flattening, or unfolding of a space-time hypercube.
Brehmer et al.~\cite{brehmer2017timelines} proposed a design space for timeline-based storytelling using layout, scale, and representation as dimensions.
Based on this design space, Timeline Storyteller~\cite{brehmer2019timeline} was designed as an authoring tool for presenting event sequence data.
Ellipsis~\cite{satyanarayan2014authoring}, another authoring tool, helps create interactive narrative visualizations as a combination of scenes, annotations, and user interactions. 
Our approach is similar in this aspect: the system automatically develops ``story pieces''~\cite{lee2015more} based on the input set of charts or visualizations (\ie, partitioning).

Data comics specifically have been identified as a form that---through the combination of familiarity, established conventions, and potential for expressive freedom---can serve as an effective medium to engage the reader, convey complexity, and enable decision making~\cite{bach2017emerging,Wang2020data}.
Bach et al.~\cite{bach2016telling} used graph-based storytelling exercises to identify design factors for creating what they call ``graph comics''.
The factors concern different aspects of graphs, their mapping to a comic-style storytelling paradigm, and their comprehension by the average viewer.

DataComicsJS~\cite{zhao2015data} is a Chrome plugin that allows users to clip existing visualizations and compose, render, and narrate them in the style of a comic-book narrator.
More recently, Datatoon~\cite{kim2019datatoon} offers a unified pen \& touch-based interactive authoring environment for analysis and presentation of networks.
Datatoon's narrative design component provides a canvas for storyboarding and composing data comics, and suggests basic comic layouts based on temporal/spatial continuity and filters.
Our work is close to Datatoon in in the application area, with the following main differences: 
(1) \name\ is an automated data comic generation system, while theirs is an authoring tool focusing more on expressivity, and 
(2) \name\ can form narratives from any set of data charts provided they are derived from the same dataset, while theirs only focuses on narration of graph-based visualizations.

\subsection{Data Story Generation and Presentation}


There has been a movement in the digital humanities to provide an orthogonal classification system that categorizes different artists and their works~\cite{bateman2017open}.
On the other end, algorithmic and graphical approaches have focused on \textit{generative} aspects, such as identifying keyframes from a continuous film to select and lay out a comic-book ``adaptation''~\cite{chu2015optimized} or simply generating, scaling, and placing text annotations within a comic~\cite{chu2013optimized}.
These approaches have not been limited to comics; computational support in the form of suggestions and refinements for have also been pursued in graphic design.
For instance, DesignScape~\cite{odonovan2015designscape} uses an energy-based model, sample styles, and design constraints (\eg, symmetry, alignment, and overlap) to suggest a set of candidate layouts given a set of graphical elements on a page.
\rv{GraphScape~\cite{kim2017graphscape} suggests a sequence of narration given a set of charts by computing ``transition cost'' based on weights computed between the charts based on their Vega-Lite specifications.
\name\ differs from GraphScape in the result: instead of a single linear sequence, \name\ uses Hullman et al.'s approach~\cite{hullman2013deeper} to generate story pieces and lays them out using Bach et al.'s design patterns~\cite{bach2018design}}.


Within narrative visualization, the interest has tended towards methodological and algorithmic approaches for layout design and caption (or annotation) generation.
Zhao \etal~\cite{zhao2015data} used elements of comics such as rendering style, characters, and captions to propose ways to generate narratives from existing visualizations.
Volder~\cite{srinivasan2019augmenting} suggests data facts to which the user can refer in order to decide on an appropriate visualization, but does not focus on the layout in storytelling.
Wang et al.~\cite{wang2019datashot} automated the generation of ``fact sheet'' by grouping data facts together into topics, and mapping the topics to graphical representations learned from a set of sample infographics.
\rv{The narrative is thus driven by the data facts and embellished by the charts.
\name\ on the other hand generates visual story pieces by grouping charts generated by analysts, and---based on the chart specifications---creates layouts and narratives that are driven by the charts and embellished by data facts.
QuickInsights~\cite{Ding2019quickinsights} provides a formulation of ``interesting patterns'' given a dataset and implements a technique to mine them.}
Calliope~\cite{Shi2020calliope} extends Wang \etal's approach using a Monte Carlo tree search algorithm to explore story pieces and present them in a logical order \rv{given tabular data as input}. 
However, they used basic predefined layout templates, based on a simple tiled layout.
\rv{In contrast to Calliope and QuickInsights, \name\ uses charts as inputs that have been created by an analyst in their data analysis process, identifies story pieces and lays them out according to Bach \etal's data comic patterns~\cite{bach2018design}.} 
Chen \etal\ proposed a workflow~\cite{chen2018supporting} for converting analysis results into a visual storytelling layout. 
Our work differs from theirs in a number of aspects: 
(1) \name{} automatically generates annotations and stitches them together, while annotations are user-generated in their case 
(2) \name{} integrates design principles for data storytelling into an automated layout generation, while theirs focuses on a basic timeline or force-directed layout;

\section{Design Rationale}

\begin{figure}[!tb]
    \includegraphics[width=\linewidth]{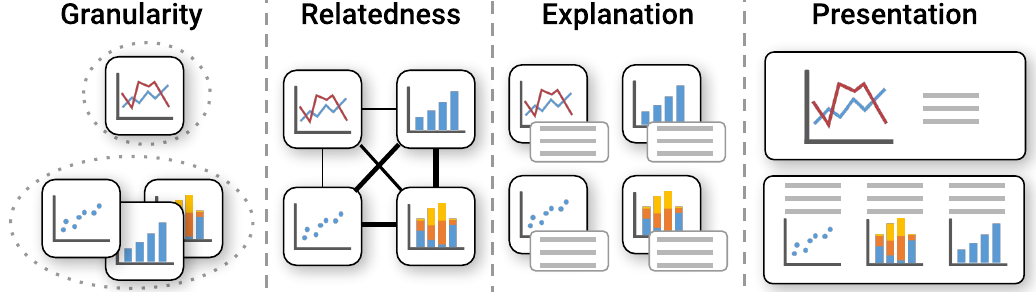}
    \vspace{-7mm}
    \caption{The \textbf{GREP} design requirements illustrated in its application to a data story generation system from a set of user-created charts.
    }
    \label{fig:grep}
\end{figure}

For \name, we base our approach on theoretical principles that have been formalized~\cite{lee2015more, hullman2013deeper, bach2018design, stolper2016emerging} for the generation, sequencing, and layout of narrative visualizations.
Based on these principles, we identify a set of requirements that we term \textbf{GREP} (\autoref{fig:grep})---\textbf{G}ranularity, \textbf{R}elatedness, \textbf{E}xplanation, and \textbf{P}resentation---for designing an automated system that generates data comics~\cite{zhao2015data} from a set of user-created charts to present a visual data story \etal\cite{lee2015more}.
In articulating these requirements, we use the term \textit{``panel''} in the same sense as a panel in a comic book, but more specifically as a combination of a \textit{single} chart with associated annotations and/or text explanation(s), contained as one unit.

\begin{description}[nosep]
   \item[\textsc{Granularity:}]
      Stories can be separated into events---or \textit{microstructures}~\cite{thorndyke1977cognitive}.
      In narrative visualization, specific facts supported by data are what form the microstructure. These facts or set of facts are called ``story pieces''~\cite{lee2015more} \rv{and would form the ``sequence of panels'' in Zhao et al.'s definition~\cite{zhao2015data} of data comics}.
        The system should identify individual story pieces based on data variables, markup, and/or visual representations extracted from a collection of charts resulted from data analysis.

   \item[\textsc{Relatedness:}]
      In contrast to microstructures, links relating events to each other form \textit{macrostructures}~\cite{thorndyke1977cognitive}.
      To create a narrative flow, connections or ``content relations''~\cite{bach2018design} need to be established \textit{within} and \textit{between} story pieces. 
      \rv{These relations would help to ``progressive development'' of the story as defined by Zhao \etal~\cite{zhao2015data}.}
      Hullman \etal~\cite{hullman2013deeper} characterize connections as ``transitions''---differences between visualizations that help in creating a linear narrative.
      These are categorized into \textit{implicit transitions}, inferrable from the data attributes/variables, and \textit{explicit transitions} that depend on the author's interpretation.
      While explicit transitions lie beyond the scope of this work, implicit transitions---such as a set of related visualizations sharing similar variables---can be automatically inferred.
        The system should characterize implicit transitions within and between story pieces in order to present a meaningful narrative. 

   \item[\textsc{Explanation:}]
      In the comic form, connections between panels are implicitly shown through a combination of ``repetition, variation, and contrast''~\cite[p.\ 12]{Groensteen2013comics}.
      Other connections in the panels are explicitly shown using text, in the form of dialogues or descriptions.
      For data comics, the narrative can be conveyed \textit{implicitly} by showing changes along one data variable while preserving overlap between panels\rv{---the ``progressive'' component in Zhao et al.'s definition~\cite{zhao2015data}}---or \textit{explicitly} by generating text explanations that supplement the visualizations~\cite{hullman2013deeper}\rv{---the ``textual elements'' in the definition}.
        The system should provide sufficient explanation of the narrative both implicitly and explicitly.

   \item[\textsc{Presentation:}]
      Panel layout is as important to creating a narrative in data comics as content relation~\cite{bach2018design}.
      For instance, a linear/sequential panel layout may be best for conveying temporal changes in a narrative, while a tiled layout may convey complementary information more effectively.
      Appropriate selection of panel layout and sequencing can reduce visual complexity, structure information, and aid understanding~\cite{wang2019comparing}\rv{---the ``visual element'' in Zhao et al.'s definition~\cite{zhao2015data}}.
        The system should thus also choose a panel layout and sequence for the story pieces that best conveys the narrative.
      Since automated narrative generation systems are not perfect,
        the system should keep the human in the loop, giving them freedom to  
        reorder items within the micro and macrostructure, alter the presentation style, and even tweak the data comic generation parameters.
\end{description}

These requirements stem from the idea of a story being built as a series of narrative tasks, such as (1) identify and group charts that are similar across multiple attributes except one, to form granular (\textbf{G}) transitions~\cite{hullman2013deeper}, (2) show relations (\textbf{R}) between charts by comparing and contrasting~\cite{bach2018design}, (3) generate text explanations (\textbf{E}) based on differences between grouped charts, and (4) use existing patterns in data comic designs~\cite{bach2018design} to present (\textbf{P}) the grouped charts.



\section{Usage Scenario}
Based on GREP, we design and develop \name{} for automating the process of crafting a data comic from a set of charts created during the exploratory data analysis.
In this section, we use a simple scenario to demonstrate the usage of the tool with a real-world dataset.

\begin{figure*}[tb]
    \centering
    \includegraphics[width=\linewidth]{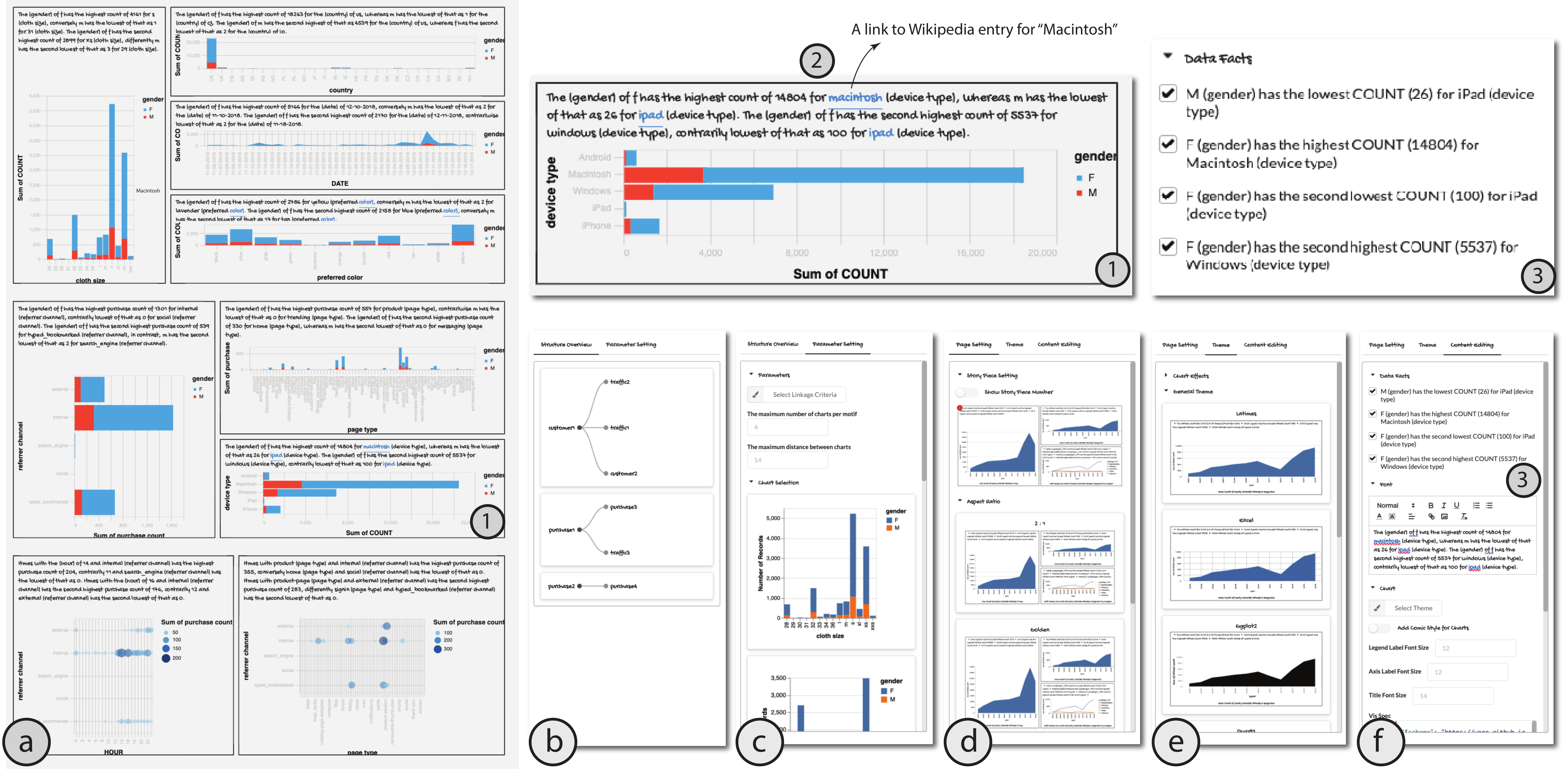}
    \vspace{-7mm}
    \caption{\name{} interface showing (a) a Data Comic Panel, an Advanced Panel with two tabs: (b) Structure Overview and (c) Parameter Setting, and a Configuration Panel with three tabs: (d) Page Setting, (e) Themes, and (f) Content Editing.
    High-res images in supplementary materials.
    }
    \label{fig:interface}
\end{figure*}

Suppose Jack, a data analyst working for an online retain store, needs to explore a dataset of the store's online visitor records and present his findings to a group of key stakeholders.
Jack conducts an exploratory visual analysis on the dataset and creates several insightful charts using an analytics software.
However, he is worried about presenting his findings in a compelling, cogent, and elegant narrative.
Composing such a narrative needs considerations of charts’ layout, order, appearance, as well as many other aspects, but the analytics software does not offer this function and Jack has little background in graphics design.
While there exists some authoring tools for creating data stories, unfortunately, he needs to present his results by the end of the workday and does not have the time to learn these tools.

Thus, Jack imports his charts into \name{}. Within seconds, a comic-strip style data story is automatically generated for him (\autoref{fig:interface}a).
He observes that \name{} identifies three coherent topics based on his findings in the charts, which are organized into three tiers (\ie, a collection of comic panels)~\cite[p.\ 50]{Chapman2012drawing}.

He sees that the tiers and layouts mostly make sense.
The first tier mainly shows observations around customer profiles.
The main chart tells about the distribution of customer visits on different \texttt{cloth sizes}, and three smaller charts display three other perspectives about the distribution of customer visits, \ie, visits by \texttt{preferred color}, \texttt{date}, and \texttt{country}.
Both the second and the third tiers describe insights around customers' product purchases.
The two charts of the second tier show the distribution of the purchases at two different aspects, \texttt{referrer channel} and \texttt{page type}, whereas the two charts of the third tier indicate more details about those purchases by \texttt{referrer channel} in terms of two dimensions, \texttt{hour} and \texttt{page type}.
However, for the first tier, Jack wants to present the last chart of the three equal-sized smaller charts first, because it indicates the fact that compared to other countries, the U.S. has the most customer visits, which is the first point he wants to present to the stakeholders about customer files.
Jack is able to easily switch these two charts in \name{}.

Jack also notices that \name{} automatically generates captions for each chart and seamlessly integrates them into the panel layout.
The text is mostly based on interesting facts observed from the underlying data of the charts, along with some contextual information for the terms.
He browses the text, and uses a side panel to further add or remove the facts for a few charts, via a list providing all the available facts (\autoref{fig:interface}f).
Then, \name{} automatically updates the corresponding captions, a natural language paragraph describing the chosen facts.
After completing the adjustment of layouts and contents, he focuses on refining the presentation styles, such as aspect ratio and font (\autoref{fig:interface}d, e).

With the compelling data story, Jack gives the following presentation:
\emph{``Looking at the first tier, the female customers are obviously more active over the male customers and the US has the most customer visits.
In detail, our customers have preferences on small-sized clothes and clothes with bright colors such as yellow and some dark colors such as black, blue, and grey.
The number of our customer visits has an outburst around two weeks before Christmas.
In terms of customers’ purchase counts, we can see that most purchases are referred from two major referrer channels.
Our customers tend to make purchases from the product pages, the home page, and the cart page.
There is also an unexpected temporal pattern that customers from one of the major channels tend to make purchases during the evening while those from the other channel commonly checkout very late at night.''}

\section{\name{} System}
\label{sec:system}
The above scenario is made possible in \name\ through a series of operations (\autoref{fig:overview}). That includes \emph{partitioning}: identification of story pieces from the overall narrative, \emph{layout}: arrangement of charts within story pieces and ordering of story pieces themselves, \emph{captioning}: generation of explanations based data facts, and \emph{refining}: stylization of the overall visual theme and adjustment of other aspects.
Please refer to our supplementary materials for more information.

\subsection{Partitioning: Identify Story Pieces}
\label{sec:identifying_story_pieces}
As previously mentioned, stories exhibit levels of granularity and are composed of \emph{microstructures} (\textbf{G}).
From a collection of charts created by a user, story pieces---subsets of charts semantically coherent and similar to each other (\eg, charts expressing similar facts and ideas)---are identified in \name{} as the first step.

\begin{figure}[!tb]
   \centering
   \includegraphics[width=\linewidth]{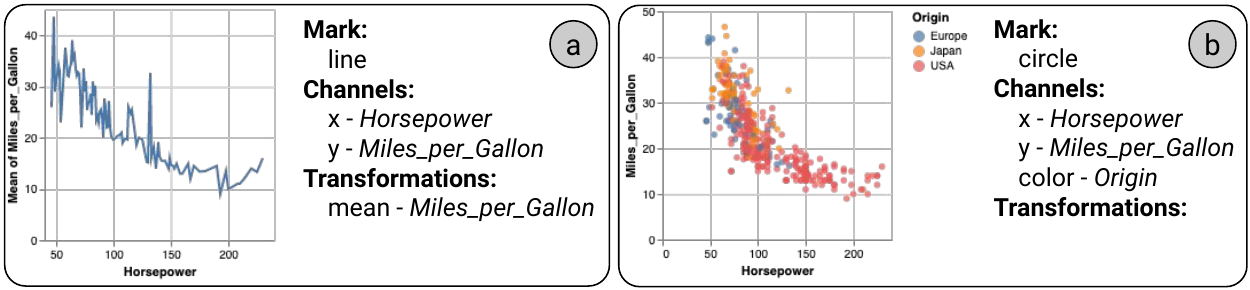}
    \vspace{-7mm}
   \caption{Characterizing charts: properties that are common and different between two charts are used to measure their distance and identify implicit story transitions. 
   }
   \label{fig:transition}
\end{figure}

To characterize the coherence between the charts, we define a metric that measures the distance between a pair of charts based on prior work~\cite{kim2017graphscape, xu2018chart, satyanarayan2017vega, moritz2019formalizing}.
Specifically, motivated by VegaLite~\cite{satyanarayan2017vega} and GraphScape~\cite{kim2017graphscape}, we represent individual charts as a set of specifications: \emph{Marks} (\eg, bar, point), \emph{Channels} (x-position, y-position, color, size), and \emph{Transformations} (\eg, sort, aggregate).
For example, \autoref{fig:transition} shows two charts characterized with these three aspects.
As the input to \name{}, we assume that the specifications are meta-data information associated with the charts.
If charts are not created with VegaLite, methods on reverse engineering charts into the VegaLite can be applied~\cite{poco2017reverse}. 

The distance between two charts is quantified by calculating the sum cost for a set of operations to ``transition'' between the two charts' specifications, based on our prior work~\cite{xu2018chart}. Considered operations include \emph{Add}, \emph{Modify}, and \emph{Remove}.
In \autoref{fig:transition}, to transition from the left chart to the right chart, the mark type is \emph{modified} from line to circle, a color channel is \emph{added}, and a mean aggregate transformation is \emph{removed}.
Each operation is assigned a numerical cost value using on the results of Kim \etal's empirical study~\cite{kim2017graphscape}.
They also introduced a directed graph model of the chart design space, GraphScape, in which nodes represent chart specifications and edges represent the operations with weight denoting the cost.
Thus, the total transition cost (\ie, distance) between two charts is computed by summing the edge weights along the shortest path traversal from one chart to another.
Unfortunately, GraphScape for real-time calculation is quite expensive (and non-interactive) due to running a breath-first search for finding the shortest path in a large graph (\ie, the chart design space). 
To achieve an interactive distance calculation, we add together the individual operation costs in light of our previous work~\cite{xu2018chart}, which takes linear time.  

Based on the calculated pairwise distances, hierarchical agglomerative clustering (HAC)~\cite{han2011data} is used to separate the charts into groups or clusters, which we consider to be the identified story pieces.
In the \name{} prototype system, we limit the maximum cluster size to four in accordance with traditional comic book design patterns~\cite{McCloud1994}, where the four-panel comic format is quite popular~\cite{fourpanel}.
We also ground this approach in patterns identified by Bach \etal~\cite{bach2018design}, whereby individual story pieces tend to contain few elements for the sake of readability and simplicity.
\name{} uses average-linkage metric to measure cluster distance, though note that other metrics, thresholds, and clustering techniques can easily be adopted as future work.

\subsection{Layout: Organize and Order Story Pieces}

\begin{figure}[tb!]
    \centering
    \includegraphics[width=\linewidth]{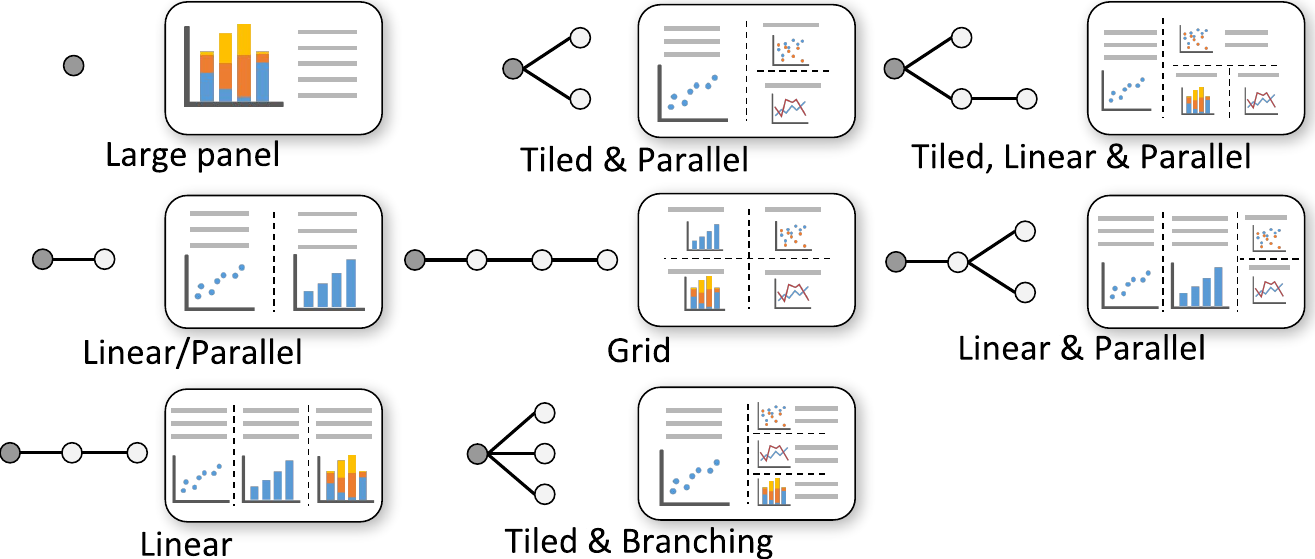}
    \vspace{-7mm}
    \caption{Matching story backbones to Bach \etal's layout patterns~\cite{bach2018design}. 
    }
    \label{fig:layout}
\end{figure}

While the previous step groups charts into distinct story pieces, the charts within each piece do not \textit{a priori} contain any inherent ordering or organization.
To present the charts with a narrative like comic strips, \name\ automatically organizes them in a two-dimensional layout (\autoref{fig:layout}) according to the connection and relation of their contents~(\textbf{R}).

Within each story piece, we construct a weighted graph of the charts with edges indicating the distances, inspired by Hullman \etal's graph-driven approach for understanding visual narratives~\cite{hullman2013deeper}.
Since every pair of charts in the graph has a calculated distance, these are complete graphs.
For each graph, we compute a minimum spanning tree (MST) to find the subsets of edges that connect all the charts with the smallest total cost.
We set the root node of each MST as the simplest chart (\ie, with the fewest specifications), based on the assumption that these charts often represent the most high-level, coarse-grained, and/or initial explorations. 
In this case, a walk-through from the root node to a leaf node is likely follow the Visual Information-Seeking Mantra~\cite{shneiderman2003eyes}, starting from an overview of data and driving down to details.
If multiple the charts have the same specification complexity, we select the earliest-created chart, assuming that users in general explore the data progressively. 
We also fine tune the MST by prioritizing charts with the same set of attributes on the paths from the root to leaf nodes.

Our layout method is grounded by the approach of maintaining consistency in visualization sequences with an objective function (minimizing the state transition cost) applied on a weighted graph of charts~\cite{hullman2013deeper}.
This MST---which we call the \emph{story backbone}---represents a narrative structure such that one can present a story piece with the highest possible continuity by following the edges.
The edges themselves represent \emph{implicit transitions} (\textbf{R}) between charts.
Based on the distance measure, charts with similar encodings and shared attributes will be more likely connected in the story backbone, thus easing the transition in storytelling.
This addresses the \textit{implicit narrative} requirement (\textbf{E}).

There are eight different combinations of story backbones for a story piece with four charts or fewer.
We map story backbones to Bach \etal's design patterns for data comic layout~\cite{bach2018design} in the presentation space (\textbf{P}).
Each story piece is laid out using one of these design patterns as a tier.
One story backbone might be reasonably matched with multiple layouts.
Currently, however, there are no agreed rules or principles for an automated matching.
Thus, based on Wang \etal's survey for layouts in infographics~\cite{wang2019datashot}, we adopt frequently-used layouts such as \emph{Tiled} (40.4\%) and \emph{Parallel} (28.6\%), and leave out rarely-used ones such as \emph{Network} (2.0\%) and \emph{Annotated} (2.9\%).
We also consider space efficiency of the layouts, \eg, avoiding too small charts and increasing the data-ink ratio~\cite{tufte1983}.
\autoref{fig:layout} shows our mappings between story backbones and layouts.

Similarly, between story pieces, we build a weighted graph using the root charts from each story piece, and then find the shortest path that connects them.
We linearly order story pieces based on this shortest path, forming the complete story that reflects its \emph{macrostructure} (\textbf{G}).

\subsection{Captioning: Generate and Integrate Explanations}
\label{sec:generating_integrating_explanations}

As seen in infographics and comic strips, a compelling story requires not only an \emph{implicit narrative} in the form of layout and sequencing, but also an \emph{explicit narrative} in the form of text explanations (\textbf{E}).
\name{} automatically generates text that describes insights for a chart, directly from the chart's data.
Based on Srinivasan \etal's work~\cite{srinivasan2019augmenting}, we can extract a large set of data facts for each chart.
For example, a data fact ``The Car with a horsepower of 193 bhp has the lowest \texttt{Miles\_per\_Gallon} (9)'' can be extracted from \autoref{fig:transition}a. 
The data fact is characterized with meta-properties including \emph{fact form}: \texttt{minimum}, \emph{fact level}: \texttt{1}, and \emph{data attributes}: \texttt{Horsepower}, \texttt{Miles\_per\_Gallon}, that are defined in \cite{srinivasan2019augmenting}.
The challenge then is to select an appropriate subset of data facts to support the explicit narrative.

Our first improvement is to propose a ranking algorithm of data facts based on the coordination between charts.
For a chart with a list of data facts $C^s=[f_0^s,f_1^s,\cdots,f_n^s]$, the weight of a data fact $f_i^s \in C^s$ is defined as:
\begin{equation}
    w_i^s =\sum_{C^t \in N(C^s)} \frac{1}{|C^t|}\sum_{f_k^t \in C^t} \alpha |F_k^t-F_i^s| + \beta |L_k^t-L_i^s| + \gamma J(A_k^t,A_i^s), \label{eq:df}
\end{equation}
Here, $N(\cdot)$ represents the neighboring charts in the story backbone;
$F$, $L$, and $A$ are the data fact meta-properties: form (\eg, correlation), level (\ie, level 1, 2, or 3), and data attributes of a data fact; 
$J(\cdot)$ is the Jaccard distance; 
and $\alpha$, $\beta$, and $\gamma$ are parameters that we set equally in our implementation.
Intuitively, this algorithm measures how much a data fact for a chart is related to all the data facts in the neighboring charts within the story piece.
If a data fact for a chart tends to have the same type, level, and data attributes as those for other charts, it will be ranked higher.
Based on the initial ranked lists of data facts, we discount duplicated facts that appear in later charts in a story piece. 

\begin{table}[tb!]
  \setlength{\extrarowsep}{2pt}
  \footnotesize
  \centering
  \caption{Examples of language stitching.} \label{tab:NLG_examples}
  \vspace{-4mm}
  \begin{tabular}{p{3.3in}}
    \hline
    \cellcolor{black!10}
    \textbf{Stitching Pattern (1): Coreference} \\ 
    \hline
    \textbf{Bulleted:} 
    $\bullet$ Explosives/ Bombs/ Dynamite (weaptype1) has the highest Attack Count (938) for 2015 (iyear). 
    $\bullet$ Explosives/ Bombs/ Dynamite (weaptype1) has the second highest Attack Count (840) for 2014 (iyear). \\
    \textbf{Stitched:}  
    The weapon type of explosives/ bombs/ dynamite has the highest attack count of 938 for the year of 2015 and also has the second highest attack count of 840 for the year of 2014. 
    \\
    \hline

    \cellcolor{black!10}
    \textbf{Stitching Pattern (2): Subordination} \\ 
    \hline
    \textbf{Bullet:} 
    $\bullet$ The Attack Count for Private Citizens \& Property is 328.67 times of that for Tourists; 
    $\bullet$ Private Citizens \& Property (targtype1) has the highest Attack Count (989). \\
    \textbf{Stitched:}  
    The attack count for private citizens \& property, which is 989 as the highest, is 328.67 times that for tourists. 
    \\
    \hline

    \cellcolor{black!10}
    \textbf{Stitching Pattern (3): Conjunction} \\ 
    \hline
    \textbf{Bullet:} 
    $\bullet$ 2004 (iyear) has the lowest Attack Count (319); 
    $\bullet$ 2014 (iyear) has the highest Attack Count (3925). \\
    \textbf{Stitched:}  
    The year of 2014 has the highest attack count of 3925, in contrast, 2004 has the lowest of that as 319. 
    \\
    \hline
  \end{tabular}
\end{table}

Our second improvement is to concatenate the top four data facts for each chart into a single text explanation that reads like a natural narrative. 
To do this, we employ three language stitching patterns commonly used in natural language generation~\cite{qian2020CHIEA}: (1) \textit{co-reference}, which eliminates subjects that are repeatedly mentioned in multiple data facts, (2) \textit{subordination} (e.g., ``which'', ``that''), which links data facts that are dependent of each other, and (3) \textit{conjunction} (e.g., ``while'', ``however''), which establishes a correlative or contrast relationships between data facts.
This concatenated explanation is then integrated into the layout generated in the previous step for presentation (\textbf{P}). 
\autoref{fig:interface}\circled{1} and \circled{3} show a comparison between the original data facts and stitched explanations; \autoref{tab:NLG_examples} shows more examples.
We also retrieve relevant information on specific terms from Wikipedia to provide context and aid understanding, accessible via a hyperlink on the term (\autoref{fig:interface}\circled{2}).

\subsection{Refining: Edit and Style Stories}

Our goal is not to completely replace human effort with automation, but to reduce the effort toward visual storytelling by pruning the design space and recommending aspects of data story creation.
While \name{} automatically generates a data comic from a collection of input charts,
it also supports a set of lightweight interactions for fine tuning the presentation of the final data comic (\textbf{P}).

These interactions include a number of standard presentation and styling options, which are accessible via a Configuration Panel with three tabs (\autoref{fig:interface}d--f). 
It can be used to control page settings such as aspect ratio (\autoref{fig:interface}d), chart stylings such as light/dark visual appearance and various themes for charts like Excel and ggplot2 (\autoref{fig:interface}e), and text content such as font settings (\autoref{fig:interface}f).


The user can change the structure and content of the data comic if not completely satisfied with the generated results.
Charts can be swapped within and between story pieces.
When this happens, the text explanations (\ie, concatenated data facts) of charts are updated automatically.
At the macrostructure level, story pieces themselves can be reordered as well.
The user can choose data facts other than the displayed ones by choosing from a ranked list shown in the second tab on the Configuration Panel (\autoref{fig:interface}d).
The data fact text can also be directly edited in place.

Finally, \name{} provides an Advanced Panel with two tabs (\autoref{fig:interface}b, c) that shows the story backbones as trees, allowing for better understanding of the data comic generation process.
The first tab (\autoref{fig:interface}b) allows a user to add or remove charts for generating the data comic (\ie, directly and indirectly editing the storybones).
The second tab (\autoref{fig:interface}c) allows an expert user to tune the parameters (\eg, weights and thresholds) of the generation process for specific goals (\eg, increasing $\beta$ in Eq.~\ref{eq:df} to have more level-consistent explanations).
To reduce the complexity, we expect this panel to be hidden for most use cases.

\section{Evaluation}

To assess \name, we characterize a set of four research questions that correspond to \name's partitioning, layout, captioning, and styling operations.

\begin{enumerate}[nosep] 
    \item[\textbf{Q1.}] \textbf{Partitioning:} Given an ensemble of charts:
    (a) does \name\ automatically cluster charts in a way that helps the creator identify story pieces, and
    (b) does each story piece convey a narrative meaning to the reader?
    \item[\textbf{Q2.}] \textbf{Layout:} Does \name\ effectively organize charts within story pieces (and the story pieces themselves) into a comic-style layout such that:
    (a) the creator needs little or no reorganization to convey the intended narrative, and
    (b) the reader can make sense of the narrative flow within and between story pieces?
    \item[\textbf{Q3.}] \textbf{Captioning:} Within each story piece, do the generated and integrated explanations for each chart:
    (a) help the creator in providing explanations to each panel and/or story piece, and
    (b) help the reader understand the narrative and context of each story piece?
    \item[\textbf{Q4.}] \textbf{Styling:} Does \name\ allow the user to customize the appearance of the data comic such that
    (a) the creator can convey their intended theme through the styling, and
    (b) the theme is understood by the reader? 
\end{enumerate}

\noindent Using \textbf{Q1--Q4}, we conduct a trio of evaluations. \textbf{Study~\#1} is a controlled user study that investigates the \textit{creator}'s perspective on the \textit{authoring} aspect, by assessing the required effort and the resultant satisfaction when creating data comics. 
In contrast, \textbf{Study~\#2} investigates the \textit{consumer}'s perspective on the \textit{readability} aspect, by assessing if \name's automated pipeline provides results comparable to what can be done manually. 
Results from these two studies show several strengths of \name{}'s automated pipeline, though the \emph{captioning} module (which in Study~\#1 originally generated data facts only in a list format without stitching) was notably criticized. 
This led to the second improvement on Srinivasan \etal's work~\cite{srinivasan2019augmenting} described in \autoref{sec:generating_integrating_explanations} of translating discrete data facts into a natural paragraph based on the stitching patterns. 
Finally, \textbf{Study~\#3} provides an assessment of the \textit{overall use} of \name{} based on a pair of interview studies with data scientists who create and review data comics using their own datasets (see the online supplementary materials for additional information).

To provide a reasonable comparison of \name{}'s automated operations in Studies~\#1 and~\#2, we developed a manual version of \name{} called \textit{Baseline}. Baseline provides access to the same set of charts and data facts as \name{}, but omits the automated backend pipeline (\autoref{fig:overview}b-d). 
To create a data comic, the user manually performs all authoring steps: placing charts into story pieces, organizing charts into a layout within each story piece, and ordering story pieces. 
Provided data facts are ordered by level only (default approach in \cite{srinivasan2019augmenting}), as opposed to the ranking (by the story piece structure) and stitching procedure outlined in \autoref{sec:generating_integrating_explanations}. 
While graphics editors such as Adobe Illustrator could serve as a baseline, they have a steep learning curve and have significantly different interfaces.

\subsection{Study~\#1: Evaluation with Data Comic Creators}

Study~\#1 is designed for data comic creators, to measure the user experience of authoring data comics with \name{}. 
As mentioned earlier, this study simulates ``handoff'' in collaborative analysis~\cite{xu2018chart, Zhao2017supporting}, where the person presenting the results is not necessarily the person(s) who conducted the analysis.
We formalized the study as a data comic-creation task for participants, using either \name{} or \baseline{}.
Measured study outcomes are \textit{task performance}, \textit{user satisfaction}, and \textit{qualitative feedback}.





\textbf{Datasets.}
We curated two study datasets based on an expert analysis of two real-world datasets: one regarding US Colleges~\cite{college_dataset} and another regarding Gun Violence~\cite{guns_dataset}.
These datasets were chosen due to their popularity on Kaggle, a data science platform that hosts over 45,000 datasets.
For each dataset, we invited a data analyst to conduct a freeform visual exploration using Jupyter Notebook~\cite{kluyver2016jupyter} and generate charts to document interesting findings.
The data analyst is a researcher who has three years of experience in both data analysis and visualization. She conducted the exploratory analysis of the datasets for the purposes of her research.
We obtained two ensembles of charts ($11$ charts for the Gun dataset, and $10$ charts for the College dataset), which are rendered using Vega-Lite~\cite{satyanarayan2017vega}. 
To ensure that visual complexity was not too high, charts were restricted to visualizing a maximum of three data attributes. 

\textbf{Participants and Apparatus.}
Twelve participants were recruited (eight males, four females), aged 20 to 28 ($\mu=23.3$, $\sigma=2.22$); all were university students (two undergraduates and $10$ graduate students) in engineering and science programs, with data analytics skills as an essential part of their training. 
Using a $7-$point Likert scale, participants self-reported a high degree of comfort in performing data analysis ($\mu=5.25$, $\sigma=1.05$) and in reading charts and data visualizations ($\mu=5.92$, $\sigma=1.00$), and a very high familiarity with comics ($\mu=6.33$, $\sigma=0.89$).
The study was conducted on a 27-inch iMac computer (3.7GHz 6-core processor, 32GB RAM, $5120\times2880$ display resolution) 
using the Chrome web browser in full-screen mode.

\textbf{Task and Design.}
We employed a within-subjects design with two independent variables: \textit{interface} (\name{} and Baseline) and \textit{dataset} (College and Guns, see below), counterbalanced using a $2\times2$ Latin square design.
The overarching task was to create a data comic using a given ensemble of charts 
and one of the interfaces.
This task was broken up into a set of five discrete subtasks \textbf{S1--S5} (\autoref{tab:task_descriptions}).
Note that the subtask requirements differ slightly between the two interfaces: Baseline requires fully manual composition, while \name\ initially automates subtasks \textbf{S2--S4} and allows subsequent manual refinement.

\begin{table}[tb!]
  \setlength{\extrarowsep}{2pt}
  \footnotesize
  \centering
  \caption{Subtask descriptions and relevant research question for Study 1.} \label{tab:task_descriptions}
  \vspace{-4mm}
  \begin{tabular}{p{3.3in}}
    \hline
    \cellcolor{black!10}
    \textbf{S1: Partitioning (Q1)} \\ 
    \hline
    \textbf{\baseline:} Divide the charts into coherent narrative story pieces of less than four charts. \\ 
    \textbf{\name:} Briefly describe the narrative of each story piece. \\
    \hline

    \cellcolor{black!10}
    \textbf{S2:~Chart~Layout (Q2)} \\
    \hline
    \textbf{\baseline:} Choose a layout from the templates for each story piece. \\
    \textbf{\name:} If necessary, rearrange the layout of each story piece. \\
    \hline

    \cellcolor{black!10}
    \textbf{S3:~Captioning (Q3)} \\
    \hline
    \textbf{\baseline:} Select less than three data facts per chart from the list. \\
    \textbf{\name:} Decide if the data facts serve the narrative; change them if necessary. \\
    \hline

    \cellcolor{black!10}
    \textbf{S4:~Story~Piece Ordering (Q2)} \\
    \hline
    \textbf{\baseline:} Order story pieces to form a narrative overall. \\
    \textbf{\name:} If necessary, reorder story pieces to form a narrative overall. \\
    \hline

    \cellcolor{black!10}
    \textbf{S5:~Styling (Q4)} \\
    \hline
    \textbf{Both:} Choose a proper style to personalize the data comic. \\
    \hline
  \end{tabular}
\end{table}

To take the study, participants created two data comics: one using \name\ and one using the Baseline.
For each interface, participants began with a training stage via a short tutorial demonstrating the interface's functionality. 
Then, \textbf{S1--5} were performed on a small testing dataset (an ensemble of eight charts from the Global Terrorism dataset~\cite{terrorism_dataset}) to acclimate participants to creating data comics using the interface.
After the training stage, the chosen dataset for the interface was loaded and \textbf{S1--5} were performed again. 
Next, participants filled out a short questionnaire that included NASA's TLX~\cite{Hart1988development} to assess the data comic creation process with that interface.
During the study, the experimenter sat beside participants to help with any confusion. 
After completing both interfaces, participants were invited to a short semi-structured interview for comparing the two interfaces on aspects such as ease of use and effectiveness.



\subsection{Results and Analysis of Study~\#1}

On average, the study took approximately 40 minutes for each participant to complete.
As Study~\#1 participants did not create the charts, thy needed time to familiarize themselves before beginning \textbf{S1} (reviewing time was also occasionally needed in subsequent subtasks). Because of this, \textit{task completion time} is not analyzed as an outcome in Study~\#1. See \autoref{sec:limitations} for more discussion.



To evaluate \name{} in Study~\#1, we analyze three primary outcomes: task performance, user satisfaction, and qualitative feedback. At a high level, they indicate \name{} effectively supports for creating data comics in a semi-automated manner, as compared to Baseline's fully manual approach. Due to a data storage issue, log and recording files for two participants were corrupted in this analysis. 

\subsubsection{Task Performance}

To analyze the task performance of \name{}, we consider data points that can act as proxies for \textbf{Q1--Q3}. As users specify styling manually both in \name{} and \baseline{}, task performance is not analyzed for \textbf{Q4}.

To investigate \textbf{Q1}, we compare how similar chart groups are between \name{} and \baseline{}.
In \name, charts could not be moved between story pieces, but \baseline{} allowed users to place charts freely within any story piece.
A high similarity between the two interfaces would indicate that \name{} partitions charts within story pieces similar to what a user would do when allowed unrestricted placement.
We compute grouping similarity using Normalized Mutual Information (NMI)~\cite{lancichinetti2009detecting}, a well-known metric for comparing clustering algorithms with ground truth labels.
However, no ground truth is available in our case, as there is no absolute correct partitioning. 
We thus use the clusters of \baseline{} as the ``ground truth'' to assess how much those of \name{} differ.  
The average NMI of \name{}, compared to \baseline{}, is $0.61$ ($\sigma=0.14$), ranged in $[0,1]$ where $1$ represents the same clustering.
It indicates \name{} generated relatively similar chart groups as the manual approach using \baseline{}. 

\begin{figure}[t]
    \centering
    \includegraphics[width=\linewidth]{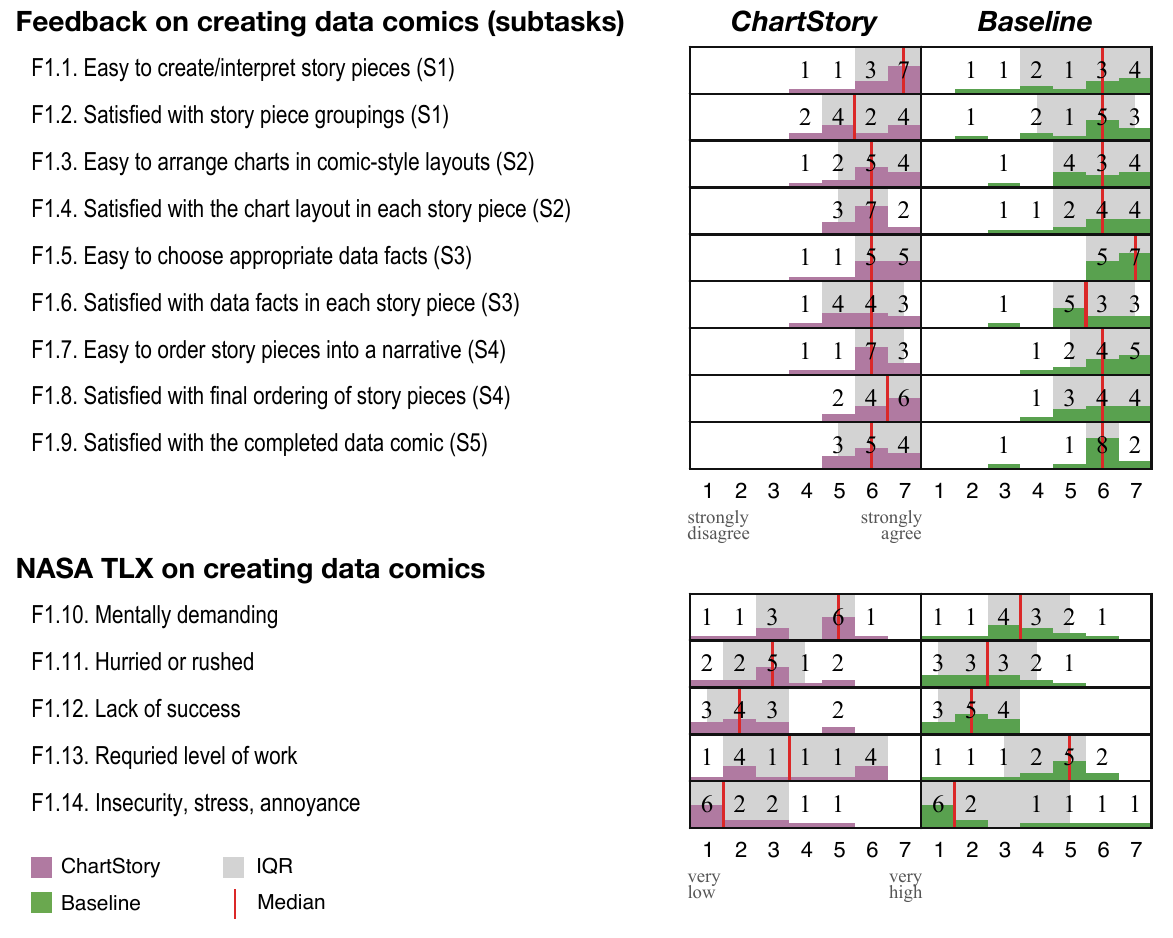}
    \vspace{-8mm}
    \caption{Participants' responses about the subtasks (F1.1--.9, the higher the better) and the overall effort (F1.10--.14, the lower the better). 
    }
    \label{fig:survey_ratings}
\end{figure}

For \textbf{Q2}, we review the number of layout edits (\ie, changing the chart layout) made to the generated story pieces of \name. On average, participants made $1.2$ ($\sigma=0.71$) layout edits per story piece, with $3$ and $4$ story pieces in total for the Gun and College datasets, respectively.
This indicates that they were generally satisfied with the layouts generated by \name{} (\autoref{fig:survey_ratings}, \textbf{F1.4}). 

To answer \textbf{Q3}, we similarly review the number of data fact edits made to \name's automatically-generated data facts. On average, users made $8.8$ ($\sigma=6.5$) data fact edits, out of more than 40 data facts provided by the system. 
The edits were participants altering the default chosen data facts (\ie, the top 4) to display on a chart, where replacing 1 data fact was counted for 2 edits: 1 for removal and 1 for addition. They did not make any text edits to the data facts. 
This likewise reflects that participants were generally satisfied with the suggested data facts (\autoref{fig:survey_ratings}, \textbf{F1.6})

While each of these data points does not in a vacuum demonstrate \name{}'s viability, they provide initial evidence that the system generates effective automated results at partitioning (\textbf{Q1}), chart layout (\textbf{Q2}), and generation of data facts (\textbf{Q3}). Further investigation of user satisfaction and qualitative feedback corroborates these findings.

\subsubsection{User Satisfaction}


User satisfaction of the data comic creation process is analyzed using the post-study questionnaire;
\autoref{fig:survey_ratings} summarizes the responses. 
We follow Dragicevic \etal's principles~\cite{dragicevic2016fair} to report the results 
in histograms rather than using p-values for accuracy and transparency. 
In the context of assessing \textbf{Q1--Q3},
these results indicate that \name{}'s automated features provide level of user satisfaction similar to manual data comic creation while potentially saving time and effort.
This process includes: 
identifying story pieces from a collection of charts (\textbf{Q1:} \textbf{F1.1, .2}), 
generating the layouts that make narrative sense in different granularity (\textbf{Q2:} \textbf{F1.3, .4, .7, .8}), 
providing text explanations with consistent narrative and context (\textbf{Q3:} \textbf{F1.5, .6}). 
In analyzing user satisfaction related to the overall process (\textbf{Q1--Q4: F1.10--.14}), we found 
there is a tendency that they felt less temporal and physical demand, more successful, and less frustrated using \name.  

\subsubsection{Qualitative Feedback}

Qualitative feedback from participants was collected during post-study interviews. 
Based on this feedback, a set of four broad sentiments shared by participants about Study~\#1's data comic creation process.

\textbf{\name\ is easier to understand and use.} Seven participants explicitly stated \name{} was the easier interface to understand and use 
(\p{1, 3, 5, 7, 9, 13, 14}), while only four participants stated \baseline{} was easier 
(\p{4, 6, 7, 12}). 
The primary justification for \name{}'s ease is that it automates tasks and layouts. 
As \p{1} noted, ``\textit{most things are automated. I just need to review it and update}.'' This was echoed by \p{5}: ``\textit{\name\ is easy to understand because it is already laid out for you. You can understand the story from each story piece.}''

\textbf{\baseline{} provides more freedom.}
When participants preferred the \baseline{}, it was generally because it provided more freedom. 
\p{4} supported this: ``\textit{I have the freedom to make any stories that I wanted. [\baseline{}] was much more flexible.}'' 
This was echoed by \p{8}: ``\textit{you can group [charts] however you want.}'' 
Only one participant (\p{12}) liked the blank slate initially provided by \baseline{}: ``\textit{For the \baseline{}, I got a fresh page to start with... it was fresh and you could create anything as you thought}.'' 
Even for some participants who thought \name{} was easier, they felt that \baseline{}'s freedom ultimately led to more satisfied results (\p{1, 7, 14}). 
For example, \p{1} commented: ``\textit{It provided me the better results because I actually have to do everything.}.''

\textbf{Manually grouping and arranging charts in \baseline{} quickly becomes tedious.}
For several participants, ``\textit{having to do everything manually}'' (\p{3})---particularly the grouping and arranging charts---quickly became a tedious endeavour. This was summarized by \p{5}: ``\textit{[for \baseline{},] I need to create from scratch in which case I need to spend a lot of time studying the visualizations. I had a hard time figuring out which visualization should go to which story piece.}'' Similarly, \p{2} disliked that \baseline{} ``\textit{doesn't have guidance for us to build a story}.'' 

\textbf{Bulleted data facts do not flow.}
Regardless of the interface, several participants mentioned the bulleted presentation of data facts hindered the narrative flow of the captions. ``\textit{The bulleted list gave me the information I wanted, but it did not have a flow}'' (\p{7}). Similarly, ``\textit{making it more [...] like in a word document would improve readability}'' (\p{5}). While participants could freely edit the data facts into a paragraph by deleting the list bullet points and changing the text to flow, this was a tedious operation to manually perform for each chart.

\subsection{Study~\#2: Evaluation with Data Comic Consumers}

To complement Study~\#1's focus on data comic creation, Study~\#2 focues the perspective of data comic \emph{consumers}, using the data comics created in Study~\#1.

\textbf{Dataset.}
Based on the output from Study~\#1, we had a collection of $10$ data comics manually created using \baseline (five each for the College and Gun datasets).
We included the ``default'' data comics for each dataset automatically generated by \name{} (\ie, without subsequent user refinement), resulting in $12$ total data comics.
We exclude subsequent user refinement as our intent is to compare the results of \name's automatically-generated comics directly against those manually generated using \baseline.

\textbf{Participants.}
We recruited $6$ expert visualization practitioners (\e{1--6}) with an average of $5$ years of experience ($\sigma=3.45$) in data-based storytelling. 
Two were data scientists (\e{1--2}) at a large digital marketing company, whose daily jobs involve reading analytics dashboards and presenting insights to help marketers design their campaigns. 
Four were university researchers (\e{3--6}) in an Information Science department, who research domain datasets and publish academic papers to report on novel data patterns revealed in visualizations.
Using expert visualization practitioners as the participants allows us to adequately assess the quality of automatically-generated and manually-created data comics.

\textbf{Task and Design.}
Participants performed rating tasks for each dataset (chart ensemble).
We first introduced the dataset.
Then, for each of the six data comics in the dataset, participants completed a questionnaire to rate its quality in terms of partitioning, layout, captioning, and styling (\textbf{Q1--4}).
Participants also provided subjective feedback about the comic's overall design.
After all data comics were individually rated, participants ranked the set based on their overall preference. This process was then repeated for the second dataset.
Both the order of datasets and the order of data comics within each dataset was counterbalanced to mitigate learning effects.
The whole study lasted about 1 hour.


\begin{figure}[t]
    \centering
    \includegraphics[width=\linewidth]{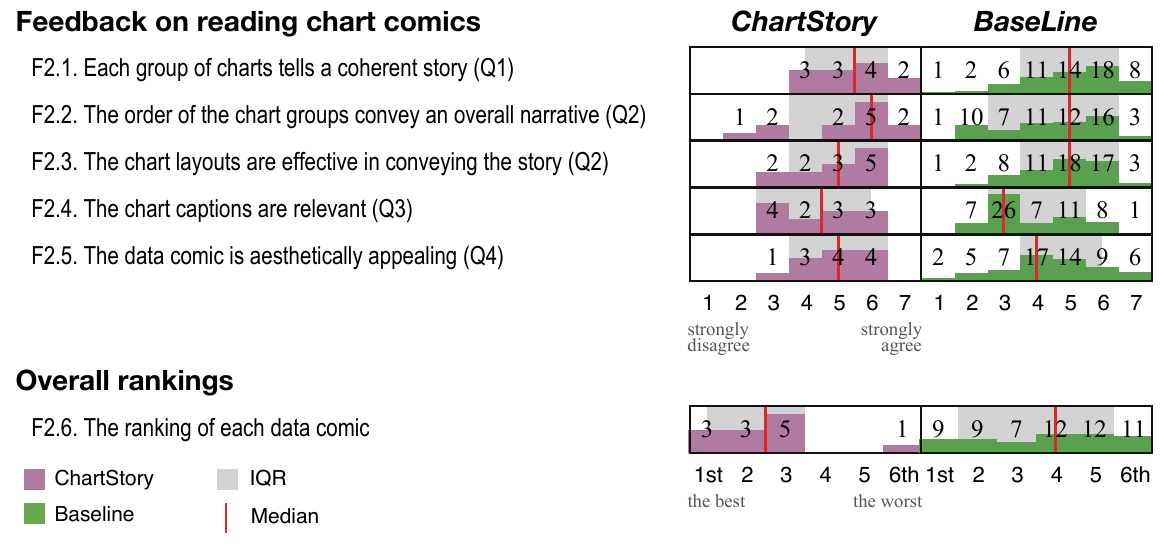}
    \vspace{-8mm}
    \caption{Participants' feedback about reading chart comics (F2.1--.5, the higher the better) and the overall ranking (F2.6, the lower the better). 
    }
    \label{fig:study2_ratings}
\end{figure}

\subsection{Results and Analysis of Study~\#2}




We report two types of results for Study~\#2: questionnaire ratings (again following Dragicevic \etal's principles~\cite{dragicevic2016fair}) and qualitative comments given about the data comics.
To provide an initial assessment, we compared the overall rankings of \name{} and \baseline{} charts (Fig~\ref{fig:study2_ratings}, \textbf{F2.6}).
On average, participants gave a higher (better) ranking for \name's automated charts ($M=2.5$, $IQR=[1.5,3]$) than the manual \baseline\ charts ($M=4$, $IQR=[2,5]$).



\subsubsection{Q1: Partitioning}
Overall, participants preferred the partitioning of \name\ ($M=5.5$, $IQR=[4.5,6]$) compared to \baseline\ ($M=5$, $IQR=[4,6]$) (\autoref{fig:study2_ratings}, \textbf{F2.1} ).
The most common issue with \baseline\ lies at its choice of variables.
For example, when reading a story about the Guns dataset, \e{3} asked: \q{Why did it use month as x-axis here but then switched to age in the next chart?} Similarly, \e{1} commented: \q{The variables are too diverse and not focused on one story... I cannot see common patterns.}
\e{1} also pointed out an issue that several story pieces in user-created comics had only one chart, and said \q{this does not tell a story at all.}

Five out of six participants thought \name\ did a better job with choosing a cohesive set of variables. \e{5}:
\q{Most of the groups have a theme.} \e{1}:
\q{Good clustering... the axes are related to each other and consistent.} \e{3}:
\q{I could see consistent x and y between charts in each story piece.}
\e{2} did not express a preference toward \name or \baseline but complained that \q{he could often find a chart not sharing any variable with others}.
One repeated suggestion (\e{1, 3, 6}) for improving \name was to combine similar story pieces. \q{The charts in story pieces 2 and 3 all have age as x-axis and should be grouped together} (\e{3}).
Only \e{4} suggested splitting a story piece: \q{I am not sure if it's good to show external factor (location) with personal factor (age) in one story piece... this misleads me to think they are correlated}.


\subsubsection{Q2: Layout}
When evaluating the order between story pieces, most participants strongly preferred the comic created using \name\ ($M=6$, $IQR=[4,6]$) over \baseline\ ($M=5$, $IQR=[3,6]$) (\autoref{fig:study2_ratings}, \textbf{F2.2}).
One observed reason for this is that the comic created using \name \q{orders the stories from simple to complex, from overview to details} (\e{3}).
\e{3} further explained that \name{}'s data comic for the College dataset \q{first reports more macro variables like month and sex, and then reports distributions over categories that are non-binary like age and place.}
Similarly, \e{1} was pleased that, \q{I can see transitions between story pieces, very good!}
Several participants proposed additional ordering rules to fit their workflow. \e{5}: \q{I prefer to always have region-related charts in the first story.} \e{4}: \q{key performance indicators such as admission rate should be presented first.}

\name\ ($M=5$, $IQR=[4,6]$) and \baseline\ ($M=5$, $IQR=[4,6]$) received similar ratings when evaluating the layout of the charts within each story (\autoref{fig:study2_ratings}, \textbf{F2.3}).
Participants observed that data comics created using \baseline\ tended to use a linear sequential layout while \name\ resulted in a more diverse structure.
\e{1} preferred \name{}'s layout because \q{the charts could be read in parallel and back-and-forth.}
He pointed at a
\baseline\ result and noted, \q{it is unclear why one chart needs to be placed after or below another.}
\e{5} also liked the
data comic created with \name: \q{It looks very compact and uses the space more efficiently.}
However, this was not always the case. \e{3} preferred \baseline\ because \q{it looks simpler, like a conventional document.}
\e{6} complained about \name: \q{The order within each story piece is unclear,} and suggested annotating the order explicitly.



\subsubsection{Q3: Captioning}
Overall, participants did not show a strong preference between the text explanations in \name\ ($M=4.5$, $IQR=[3,5.5]$) and \baseline\ ($M=3$, $IQR=[3,5]$) (\autoref{fig:study2_ratings}, F2.4).
\e{1} found text explanation helpful in both contexts, as \q{the text can make sure that people are on the same page and getting the same insights and conclusions from the same chart.}
\e{2} added that \q{when the chart shows too much data or the trends are not easy to see, captions can make me aware of them.}
Similarly, \e{4} mentioned that \q{the text loses some information but captures the key signals with explanations.}

All participants agreed that the captions can be improved.
Many found that some of the data facts were not informative and not easy to read, as \e{1} explained: \q{The caption says what is high, what is low... I can see it in the chart.}
To mitigate this issue, \e{1} and \e{4} suggested 
emphasizing data facts that \q{you could not see by eye} were not directly evident in the charts, such as the trends and comparisons: \q{Looking at the bar chart, I think the comparison between bars and the trend of growth give more insights than the counts, which help me tell a better story.}
As a suggestion for improvement, \e{6} asked for story piece-level captions (as opposed to captions appended to individual charts) that provide \q{global context about the shared key variable across the charts.}

To enhance the readability of the captions, \e{1} proposed: \q{Instead of showing a ranked list, we can group the facts by topics or order by overall to detail to tell a story, like the captions in news articles.}
Similarly, \e{2} suggested that \q{the bullet points are not super readable... I want something more natural that I can read aloud to my colleague.}


\subsubsection{Q4: Styling}
While all the data comics shared similar visual styles (chart types, mark and channel encodings, color schemes, etc.), participants found \name\ ($M=5$, $IQR=[4,6]$) more aesthetically pleasing than \baseline\ ($M=4$, $IQR=[4,5.5]$) (\autoref{fig:study2_ratings}, \textbf{F2.5}).
For example, when reviewing \baseline's data comics, \e{2} complained that some charts \q{are in different sizes and hard to compare}. At one point, \e{1} pointed at a story piece and noted: \q{The charts are overlaid by mistake... better make sure the charts are aligned and axes are aligned.} 
In contrast, none of these defects were observed with \name. \e{1, 3, 6} individually applauded that it looks polished and uniform.

Suggestions were made to improve both systems regarding the choices of visual encodings and color schemes:
\q{same variables should be encoded consistently and different variables should be encoded differently} (\e{1}),
\q{I don't like the stacked bar chart of month vs. number of records... the values are very close
} (\e{5}),
and \q{the heatmap is impossible to read... hard to differentiate green, yellow, blue, and the gradients in between} (\e{6}).

Strategies were also proposed for better utilizing the space. For example, \e{3} suggested that \q{simpler charts should get less space... the heatmap should get more space because it has more information.}
\e{4} asked to allocate more space to important story pieces, and 
\e{6} suggested that \q{similar charts should get even sizes and placed side by side... it makes the comparison easier.}
These comments underscore how \name is ideal for use by analysts who perform the analysis and want to communicate their findings.
By automating the data comic generation, \name not only provides a way of conveying the story, but also allows the analyst to reflect on aspects of their analysis/visualization that may be absent from their collection of charts.
This allows them to iterate between analysis and presentation, with very little effort spent on the presentation itself.



\subsection{Study~\#3: Interview Studies with Domain Data}

As a follow-up to Studies~\#1 and~\#2, which looked at data comic creation and consumption in discrete, controlled settings, we conducted an interview study to holistically assess \name{} in the context of real-world data applications.
For this, we coordinated with two data scientists from Study~\#2 (\e{1--2}), who work at a large digital marketing company.
The data scientists mainly use Vega and Vega-Lite as the main charting tool in their daily workflow. 
They provided a domain marketing dataset of an online retail store, along with a set of nine charts created using Voyager~\cite{wongsuphasawat2015voyager}.
Voyager was chosen because it is one of their frequently used tools and it can easily export the charts in Vega-Lite for the input to \name{}.

The marketing dataset includes $27,780$ customer profiles (e.g., \texttt{gender}, \texttt{country}, and \texttt{cloth\_size}) and their online journeys (e.g., \texttt{referrer\_channel}, \texttt{product\_views}, and \texttt{purchase\_count}). The provided charts
were made to analyze performance of the online store, visualizing the growth of the traffic and revenue, the composition of the customers, and customers' common behavioral patterns.


\name\ generated a data comic containing four story pieces from the charts.
The first story piece consisted of four barcharts, separating the number of customers by \texttt{cloth\_size}, \texttt{preferred\_color}, \texttt{gender}, and \texttt{country}, respectively.
The second showed a scatterplot of \texttt{gender} by \texttt{device\_type}.
The third showed two side-by-side barcharts of \texttt{purchase\_count} categorized by \texttt{referrer\_channel} and \texttt{product\_views}.
The last story piece used two scatterplots to further break down \texttt{referrer\_channel} and \texttt{product\_views} by \texttt{hour\_of\_day}, with the sizes of the dots encoding \texttt{purchase\_count}.

Upon seeing the generated data comic, \e{1} and \e{2} immediately remarked on the value of the automated data comic creation.
\e{1} contrasted the data comic to more linear presentation formats (such as data notebooks): \q{It is much more meaningful to look at the stories than just a bunch of charts ordered by file names or creation time.}
Interestingly, \e{2} considered the charts as a form of presentation dashboard: \q{The stories are very cohesive... good dashboard.}
After reviewing the story pieces in detail, both suggested combining the first two story pieces, since the charts are involve different characteristics of the customer profiles.
\e{2} also suggested combining story piece 3 and 4 (since they likewise share many variables), but \e{1} disagreed: \q{I prefer to keep them separate since they are at different levels of details.}
Both \e{1} and \e{2} agreed that \name\ provides a good starting point for organizing charts, requiring only minor manual refinements.
This was succinctly noted by \e{1}: \q{I am surprised that it is fully automatic... it will definitely save me and my colleagues a lot of time.}

Study~\#1 and~\#2 revealed to us that the bulleted presentation of data facts was frustrating at demonstrating a coherent story narrative.
Therefore, for this study, we employed the ``language stitching'' technique (see \autoref{sec:generating_integrating_explanations} and \autoref{fig:interface}).
Both participants, who had previously seen the bulleted presentation in Study~\#2, reported preferring the new text explanations.
\e{1} said: \q{I like the full sentence captions... it is a better form to tell a story... bullet points are less formal to read}.
\e{2} commented: \q{The language is more natural... it is closer to the captions I need in my reports and slides... this feature makes my job easier.}

At the end of the interview, both participants expressed their excitement about \name\ and asked for long-term deployment studies. They also requested additional features, including generating story-level captions, supporting filters for exploring data comics, and adding effects to make the exploration more interactive. 

\section{Discussion}

\begin{figure*}[!tb]
    \centering
    \includegraphics[width=\linewidth]{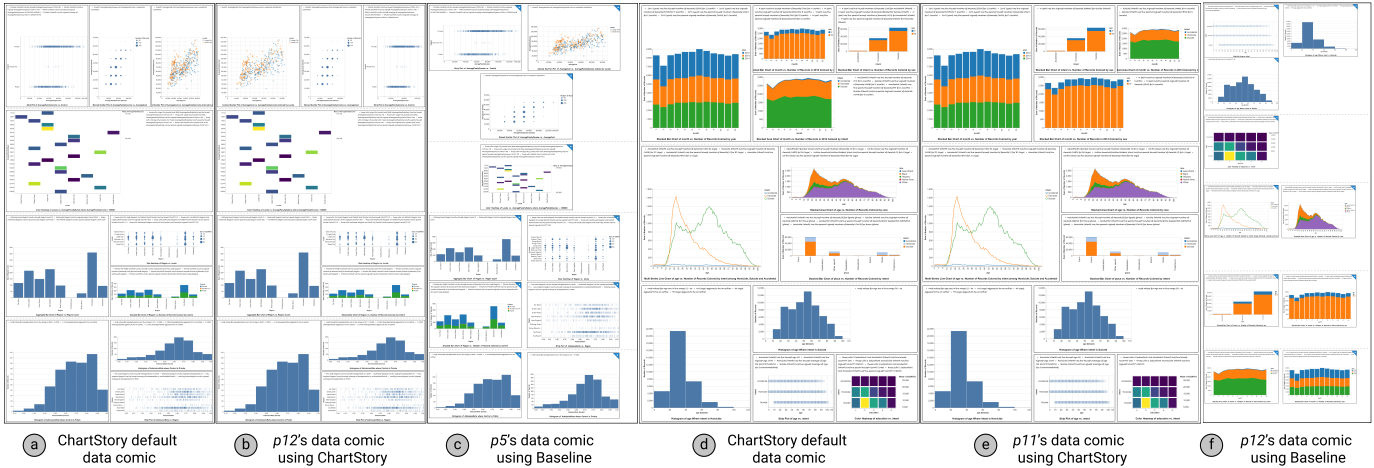}
    \vspace{-7mm}
    \caption{\rv{Data comic examples of the College (a, b, c) and the Gun (d, e, f) datasets. (a, d) The default data comics generated by \name{}. (b, e) The finished data comics by participants using \name{}. (c, f) The finished data comics by participants using \baseline{}.} 
    }
    \label{fig:example_study_charts}
\end{figure*}

Structured by the research questions \textbf{Q1}--\textbf{Q4}, we discuss the implications from our studies on  semi-automated data comic creation and consumption.

\textbf{Q1: Partitioning.} 
Study~\#1's survey responses indicated that \name{}'s solution was as high as \baseline's manual grouping, even though \name\ did not allow chart switches across story pieces (which was disabled for the study to reduce confounds).
The experts' feedback in Study~\#2 indicated this constraint was a good idea: participants' manual grouping of story pieces often went against some fundamental visualization and storytelling principles, such as maintaining continuity across plots in a story piece.
\name, with its distance-based approach to partitioning, avoids this pitfall, and allows for creation of additional metrics for a more consistent and coherent output.

\textbf{Q2: Layout.}
Using \name, participants from Study~\#1 were able to achieve an equivalent level of satisfaction with arranging charts in a significantly shorter amount of time.
They felt the layout freedom provided by \baseline\ was an advantage: ``\textit{I can do what I want to layout and organize the story}'' (\p{2}).
However, we observed that---with this freedom---all participants violated design principles for comic strips, such as arranging comics in non-space-filling layouts (with significant white space) and positioning adjacent cells in uneven alignments, \rv{even though half of them were exposed to \name{} first in the study.
\autoref{fig:example_study_charts} compares the data comics by participants using \name\ with the ones using \baseline.
The layouts in \autoref{fig:example_study_charts}-b, e is close to the initial automated output (\autoref{fig:example_study_charts}-a, d), while \autoref{fig:example_study_charts}-c, f show haphazardly-arranged charts created using \baseline.}
This difference was also noticed in Study~\#2. 
\name\ was mostly preferred because of the story ordering (simple to complex), ease of comparison between charts, compact layout, and transitions between story pieces; 
however, \baseline\ was also preferred because it followed a more conventional ``document-like'' layout.

\textbf{Q3: Captioning.}
The initial caption generation technique in Study~\#1 and~\#2 was based on Srinivasan \etal's work~\cite{srinivasan2019augmenting}, with the data facts ranked based on the relationships between charts within a story piece.
In both studies, our ranking technique did not fare differently from the original method~\cite{srinivasan2019augmenting} that is used in \baseline.
However, participants found it difficult to read captions in the form of separate bulleted items, though they acknowledged that the data facts provided a good starting point for captioning.
Our language stitching approach (\autoref{sec:generating_integrating_explanations}) was better received in our follow-up studies, who stated that the captions were closer to what they would use.
Moving forward, we plan to explore natural-language generation approaches that can synthesize higher-level captions based on data facts prioritized by the user.
This could help address the issues of low-level and high-level storytelling uncovered in the study.

\textbf{Q4: Styling.}
Participants from Study~\#1 tended to find it tedious to give the data comic a finished look using \baseline, as dragging and resizing the charts to fit the available two-dimensional layout space required manual sizing and placement. As a result, data comics created using \baseline\ looked slightly unfinished (\eg, \autoref{fig:example_study_charts}-c), in contrast to the layouts automatically generated by \name.
This difference was also noticed by experts in Study~\#2, who preferred the \q{polished and uniform} look of data comics created by \name, compared to those created using \baseline. In Study~\#2, the main criticisms of \name\ involved functionalities also shared in \baseline, such as color palette and granularity of data encoding in the charts themselves.
Control of such parameters is given to the user currently, though it is feasible to specify rules for such attributes which will further ensure uniformity and aid easy creation of data comics without sacrificing the quality and aesthetics.


\textbf{Broader Implications.} 
Besides above aspects specific to each research question, our development and evaluation of \name{} deals with several generalizable takeaways regarding human-machine collaboration.
First, the design of automated tools should focus on assisting users with tasks that they do not excel, to improve the overall productivity.
With \name, analysts can focus on creating charts (that they are good at) to better serve the story and worry less about layout or presentation (that they are not as good at). This is similar to, for example, typesetting using \LaTeX, which allows the writer to focus on content instead of document layout and styling.

Second, although auto-generated results may not be perfect---as evidenced by study users performing subsequent manual refinements in Studies~\#1 and~\#3---this approach can provide a serendipitous benefit for users by enlightening them, while improving their efficiency. As an example, in normal practice, a user has to manually write the textual annotations for a data comic, which is a time-consuming process. \name{} automatically generates text items, thus provides a good starting point for further refinement requiring much less effort to arrive at the ``final'' polished version of the data comic.

Third, there is a trade-off between user freedom and constraint in developing semi-automated tools. As discussed earlier, while our study population claimed high familiarity with comics, they showed poor skill at actually designing them. \name{} operationalizes the design patterns by Bach et al.~\cite{bach2018design} to help users achieve a good design by constraining their operations.
In comparison, \baseline\ users---in spite of having complete freedom---created sub-optimal layouts.
However, too much constraint may result in issues in trust, which is another factor to consider. Determining the appropriate balance between freedom vs. constraint is beyond the scope of the current work.


\section{Limitations and Future Work}
\label{sec:limitations}

While the study results indicate \name{} is promising to help analysts generate compelling data comics. The system and our study design still have limitations. 

First, the MST currently prioritizes charts (attribute) similarity based on the implicit transitions proposed by Hullman \etal~\cite{hullman2013deeper}.
However, visually-similar charts need not always show related data, or be part of the same narrative.
This problem could be exacerbated by scale: more charts to partition results in a higher chance of more unrelated charts clustered together.
The MST we propose is illustrative and not comprehensive, and it is a robust enough technique that can adapt to new measures of similarity or ``relatedness.''
For example, with advanced machine learning techniques, an embedding space of charts could be learned using a similar method by Zhao \etal\cite{Zhao2020chartseer}, so that the relatedness can be better captured.
This flexible notion of relatedness worth more exploration in the future.
Also, enlightened by the theories of Cohn \cite{Cohn2013}, more studies can be conducted to investigate users' strategies of reading the data comics with different panel layouts and how such relatedness plays a role in their comprehension. 
These studies will further shed light on understanding the effect of comic design layout patterns in \name{} in assisting storytelling, compared to the ``random facts'' in Wang \etal\cite{wang2019datashot}.

Second, \name\ may be limited in processing a large collection of charts, which results in a large number of story pieces. 
However, it does not make sense for an analyst to present a large number of charts in a single narrative, which overwhelms their audience and defeats the purpose of storytelling. 
Some algorithms may be developed to help select the most relevant charts before generating a data comic, or rank different data comics generated from the charts based on certain criteria.
Also, \name{} limits each story piece with a maximum of four charts, based our observations of Bach \etal's patterns~\cite{bach2018design} and the prevalence of four-panel comic~\cite{fourpanel}. 
Future studies could be conducted on investigating the effect of story piece size in \name{}.

Third, Study~\#1 suffered from the limitation that participants were not the people who performed the data analysis and created the charts in the first place.
While the study was indeed based on the existing scenario of handoff in asynchronous collaborative analysis~\cite{xu2018chart, Zhao2017supporting}, the limitation was also imposed for the sake of uniformity---allowing users to create their own charts would have resulted in the number and relevance of the charts affecting the outcome and the narrative.
By keeping the input charts uniform across participants, we mitigate any variations that may be caused by individual differences in the analysis.
Also, we introduce the issue of comprehension---some participants may be able to read and interpret the input charts quicker than others.
Thus, we did not use ``data comic creation time'' as an indicator of how the usefulness of \name.
Study~\#2 involved a relatively small number of participants, an unavoidable constraint when participants are required to be experts/practitioners.
We plan to conduct additional evaluation of \name{} to address these issues.

Lastly, we focus on a specific type of data comics defined by Zhao \etal\cite{zhao2015data}.
While the outputs of \name{} already resemble several attributes of general comic strips (\eg, panel layout and annotation), there are certain aspects missing.  
For example, comic strips usually have characters to carry out the story, speech bubbles to narrate information, and special visual effects to emphasize story elements. 
However, past studies indicate that data comics do not usually have a character~\cite{bach2018design,bach2017emerging}, which draw principles from comics and combine techniques in infographics and data visualization.
To resemble the speech bubbles, the generated explanations of \name{} can be placed in-situ on the charts when appropriate. 
It also worth exploring techniques that use special effects to emphasize certain parts of a chart, such as magnifying outliers or calling out key patterns with distortion based or picture-in-picture methods.

\section{Conclusion}
We have presented \name{}, a tool that helps analysts craft a data story in comic-strip style from charts generated in their visual exploration of data.
\name{} provides an analytical pipeline to automate the partitioning, layout, and captioning of data comics, as well as a number of intuitive interactions to allow for further refining and styling.
The design of \name{} is iteratively improved, grounded by a set of design rationale (GREP) distilled from the literature. 
We conducted a comprehensive evaluation of \name{} by comparing against a manual baseline with data comic creators and consumers, followed by in-depth interviews with two data scientists.
The results indicated that \name{} can provide cogent recommendations for creating data comics that make narrative sense to participants and compare favorably to data comics created by the baseline.


%



\ifCLASSOPTIONcompsoc
  \section*{Acknowledgments}
\else
  \section*{Acknowledgment}
\fi

This work is supported in part by the Natural Sciences and Engineering Research Council of Canada, the
U.S. National Science Foundation through grants IIS-1741536, IIS-1528203, and OAC-1934766, and an Adobe gift fund.

\ifCLASSOPTIONcaptionsoff
  \newpage
\fi



\bibliographystyle{abbrv}
%


\bibliography{chartstory}

\begin{thebibliography}{10}

\bibitem{college_dataset}
{College Scorecard Data}.
\newblock \url{https://collegescorecard.ed.gov/data}, 2019.

\bibitem{terrorism_dataset}
{Global Terrorism Database | Kaggle}.
\newblock \url{https://www.kaggle.com/START-UMD/gtd}, 2019.

\bibitem{guns_dataset}
{Gun Deaths in the US: 2012-2014 | Kaggle}.
\newblock \url{https://www.kaggle.com/hakabuk/gun-deaths-in-the-us}, 2019.

\bibitem{fourpanel}
{Why 4-Panel Comics Now Dominate Our Screens | Wired}.
\newblock \url{https://www.wired.com/story/four-panel-webcomics/}, 2019.

\bibitem{bach2017descriptive}
B.~Bach, P.~Dragicevic, D.~Archambault, C.~Hurter, and S.~Carpendale.
\newblock A descriptive framework for temporal data visualizations based on
  generalized space-time cubes.
\newblock In {\em Computer Graphics Forum}, volume~36, pages 36--61, 2017.

\bibitem{bach2016telling}
B.~Bach, N.~Kerracher, K.~W. Hall, S.~Carpendale, J.~Kennedy, and
  N.~Henry~Riche.
\newblock Telling stories about dynamic networks with graph comics.
\newblock In {\em Proc. of the ACM Conference on Human Factors in Computing
  Systems}, pages 3670--3682, 2016.

\bibitem{bach2017emerging}
B.~Bach, N.~H. Riche, S.~Carpendale, and H.~Pfister.
\newblock The emerging genre of data comics.
\newblock {\em IEEE computer graphics and applications}, 37(3):6--13, 2017.

\bibitem{bach2018design}
B.~Bach, Z.~Wang, M.~Farinella, D.~Murray-Rust, and N.~H. Riche.
\newblock Design patterns for data comics.
\newblock In {\em Proc. of the ACM Conference on Human Factors in Computing
  Systems}, 2018.

\bibitem{bateman2017open}
J.~A. Bateman, F.~O. Veloso, J.~Wildfeuer, F.~H. Cheung, and N.~S. Guo.
\newblock An open multilevel classification scheme for the visual layout of
  comics and graphic novels: Motivation and design.
\newblock {\em Digital scholarship in the humanities}, 32(3):476--510, 2017.

\bibitem{brehmer2017timelines}
M.~Brehmer, B.~Lee, B.~Bach, N.~H. Riche, and T.~Munzner.
\newblock Timelines revisited: A design space and considerations for expressive
  storytelling.
\newblock {\em IEEE Trans. on Visualization and Computer Graphics},
  23(9):2151--2164, 2017.

\bibitem{brehmer2019timeline}
M.~Brehmer, B.~Lee, N.~Henry~Riche, D.~Tittsworth, K.~Lytvynets, D.~Edge, and
  C.~White.
\newblock Timeline storyteller: The design \& deployment of an interactive
  authoring tool for expressive timeline narratives.
\newblock In {\em Proc. of the Computation+Journalism Symposium}, pages 1--5,
  2019.

\bibitem{Chapman2012drawing}
R.~Chapman.
\newblock {\em Drawing Comics Lab: 52 Exercises on Characters, Panels,
  Storytelling, Publishing \& Professional Practices}.
\newblock Quarry Books, Beverly, MA, 2012.

\bibitem{chen2018supporting}
S.~Chen, J.~Li, G.~Andrienko, N.~Andrienko, Y.~Wang, P.~H. Nguyen, and
  C.~Turkay.
\newblock Supporting story synthesis: Bridging the gap between visual analytics
  and storytelling.
\newblock {\em IEEE Trans. on visualization and computer graphics}, pages 1--1,
  2018.

\bibitem{chu2013optimized}
W.-T. Chu and C.-H. Yu.
\newblock Optimized speech balloon placement for automatic comics generation.
\newblock In {\em Proc. of the int'l workshop on Interactive multimedia on
  mobile \& portable devices}, pages 1--6, 2013.

\bibitem{chu2015optimized}
W.-T. Chu, C.-H. Yu, and H.-H. Wang.
\newblock Optimized comics-based storytelling for temporal image sequences.
\newblock {\em IEEE Trans. on Multimedia}, 17(2):201--215, 2015.

\bibitem{Cohn2013}
N.~Cohn.
\newblock Navigating comics: An empirical and theoretical approach to
  strategies of reading comic page layouts.
\newblock {\em Frontiers in Psychology}, 4:186, 2013.

\bibitem{Ding2019quickinsights}
R.~Ding, S.~Han, Y.~Xu, H.~Zhang, and D.~Zhang.
\newblock Quickinsights: Quick and automatic discovery of insights from
  multi-dimensional data.
\newblock In {\em Proceedings of the ACM International Conference on Management
  of Data}, pages 317--332, 2019.

\bibitem{dragicevic2016fair}
P.~Dragicevic.
\newblock Fair statistical communication in {HCI}.
\newblock In {\em Modern statistical methods for HCI}, pages 291--330.
  Springer, 2016.

\bibitem{Groensteen2013comics}
T.~Groensteen.
\newblock {\em Comics and narration}.
\newblock Univ. Press of Mississippi, 2013.

\bibitem{han2011data}
J.~Han, M.~Kamber, and J.~Pei.
\newblock {\em Data Mining: Concepts and Techniques}.
\newblock Morgan Kaufmann, 2011.

\bibitem{Hart1988development}
S.~G. Hart and L.~E. Staveland.
\newblock Development of nasa-tlx (task load index): Results of empirical and
  theoretical research.
\newblock In {\em Advances in psychology}, volume~52, pages 139--183. 1988.

\bibitem{hullman2013deeper}
J.~Hullman, S.~Drucker, N.~H. Riche, B.~Lee, D.~Fisher, and E.~Adar.
\newblock A deeper understanding of sequence in narrative visualization.
\newblock {\em IEEE Trans. on Visualization and Computer Graphics},
  19(12):2406--2415, 2013.

\bibitem{Kery2017exploring}
M.~B. Kery and B.~A. Myers.
\newblock Exploring exploratory programming.
\newblock In {\em Proc. of the IEEE Symposium on Visual Languages and
  Human-Centric Computing}, pages 25--29. IEEE, 2017.

\bibitem{kim2019datatoon}
N.~W. Kim, N.~Henry~Riche, B.~Bach, G.~Xu, M.~Brehmer, K.~Hinckley, M.~Pahud,
  H.~Xia, M.~J. McGuffin, and H.~Pfister.
\newblock Datatoon: Drawing dynamic network comics with pen+touch interaction.
\newblock In {\em Proc. of the ACM Conference on Human Factors in Computing
  Systems}, number 105, 2019.

\bibitem{kim2017graphscape}
Y.~Kim, K.~Wongsuphasawat, J.~Hullman, and J.~Heer.
\newblock Graphscape: A model for automated reasoning about visualization
  similarity and sequencing.
\newblock In {\em Proc. of the ACM Conference on Human Factors in Computing
  Systems}, 2017.

\bibitem{kluyver2016jupyter}
T.~Kluyver, B.~Ragan-Kelley, F.~P{\'e}rez, B.~E. Granger, M.~Bussonnier,
  J.~Frederic, K.~Kelley, J.~B. Hamrick, J.~Grout, S.~Corlay, et~al.
\newblock Jupyter notebooks-a publishing format for reproducible computational
  workflows.
\newblock In {\em ELPUB}, pages 87--90, 2016.

\bibitem{lancichinetti2009detecting}
A.~Lancichinetti, S.~Fortunato, and J.~Kert{\'e}sz.
\newblock Detecting the overlapping and hierarchical community structure in
  complex networks.
\newblock {\em New journal of physics}, 11(3):033015, 2009.

\bibitem{lee2015more}
B.~Lee, N.~H. Riche, P.~Isenberg, and S.~Carpendale.
\newblock More than telling a story: Transforming data into visually shared
  stories.
\newblock {\em IEEE Computer Graphics and Applications}, 35(5):84--90, 2015.

\bibitem{mackinlay1986automating}
J.~Mackinlay.
\newblock Automating the design of graphical presentations of relational
  information.
\newblock {\em ACM Trans. On Graphics}, 5(2):110--141, 1986.

\bibitem{McCloud1994}
S.~McCloud.
\newblock {\em Understanding Comics: The Invisible Art}.
\newblock William Morrow, 1994.

\bibitem{moritz2019formalizing}
D.~Moritz, C.~Wang, G.~L. Nelson, H.~Lin, A.~M. Smith, B.~Howe, and J.~Heer.
\newblock Formalizing visualization design knowledge as constraints: Actionable
  and extensible models in draco.
\newblock {\em {IEEE} Trans. on Visualization and Computer Graphics},
  25(1):438--448, 2019.

\bibitem{odonovan2015designscape}
P.~O'Donovan, A.~Agarwala, and A.~Hertzmann.
\newblock {DesignScape}: Design with interactive layout suggestions.
\newblock In {\em Proc. of the ACM Conference on Human Factors in Computing
  Systems}, pages 1221--1224, 2015.

\bibitem{poco2017reverse}
J.~Poco and J.~Heer.
\newblock Reverse-engineering visualizations: Recovering visual encodings from
  chart images.
\newblock In {\em Computer Graphics Forum}, volume~36, pages 353--363. Wiley
  Online Library, 2017.

\bibitem{qian2020CHIEA}
X.~Qian, E.~Koh, F.~Du, S.~Kim, and J.~Chan.
\newblock A formative study on designing accurate and natural figure captioning
  systems.
\newblock In {\em Extended Abstracts of the ACM Conference on Human Factors in
  Computing Systems}, pages 1--6, 2020.

\bibitem{ren2017chartaccent}
D.~Ren, M.~Brehmer, B.~Lee, T.~Hollerer, and E.~K. Choe.
\newblock {ChartAccent}: Annotation for data-driven storytelling.
\newblock In {\em Proc. of the Pacific Visualization Symposium}, 2017.

\bibitem{Rule2018exploration}
A.~Rule, A.~Tabard, and J.~D. Hollan.
\newblock Exploration and explanation in computational notebooks.
\newblock In {\em Proc. of the ACM Conference on Human Factors in Computing
  Systems}, pages 1--12, 2018.

\bibitem{satyanarayan2014authoring}
A.~Satyanarayan and J.~Heer.
\newblock Authoring narrative visualizations with ellipsis.
\newblock In {\em Computer Graphics Forum}, volume~33, pages 361--370, 2014.

\bibitem{satyanarayan2014lyra}
A.~Satyanarayan and J.~Heer.
\newblock Lyra: An interactive visualization design environment.
\newblock {\em Computer Graphics Forum}, 33(3):351--360, 2014.

\bibitem{satyanarayan2017vega}
A.~Satyanarayan, D.~Moritz, K.~Wongsuphasawat, and J.~Heer.
\newblock Vega-lite: A grammar of interactive graphics.
\newblock {\em {IEEE} Trans. on Visualization and Computer Graphics},
  23(1):341--350, jan 2017.

\bibitem{Shi2020calliope}
D.~{Shi}, X.~{Xu}, F.~{Sun}, Y.~{Shi}, and N.~{Cao}.
\newblock Calliope: Automatic visual data story generation from a spreadsheet.
\newblock {\em IEEE Trans. on Visualization and Computer Graphics}, pages 1--1,
  2020.

\bibitem{shneiderman2003eyes}
B.~Shneiderman.
\newblock The eyes have it: A task by data type taxonomy for information
  visualizations.
\newblock In {\em The craft of information visualization}, pages 364--371.
  Elsevier, 2003.

\bibitem{srinivasan2019augmenting}
A.~Srinivasan, S.~M. Drucker, A.~Endert, and J.~Stasko.
\newblock Augmenting visualizations with interactive data facts to facilitate
  interpretation and communication.
\newblock {\em IEEE Trans. on Visualization and Computer Graphics},
  25(1):672--681, 2019.

\bibitem{stolper2016emerging}
C.~D. Stolper, B.~Lee, N.~H. Riche, and J.~Stasko.
\newblock Emerging and recurring data-driven storytelling techniques: Analysis
  of a curated collection of recent stories.
\newblock Technical report, Microsoft Research, 2016.

\bibitem{Subramanian2020tractus}
K.~Subramanian, J.~Maas, and J.~Borchers.
\newblock Tractus: Understanding and supporting source code experimentation in
  hypothesis-driven data science.
\newblock In {\em Proc. of the ACM Conference on Human Factors in Computing
  Systems}, pages 1--12, 2020.

\bibitem{thomas2005illuminating}
J.~J. Thomas and K.~A. Cook.
\newblock Illuminating the path: The research and development agenda for visual
  analytics.
\newblock Technical report, PNNL, 2005.

\bibitem{thorndyke1977cognitive}
P.~W. Thorndyke.
\newblock Cognitive structures in comprehension and memory of narrative
  discourse.
\newblock {\em Cognitive psychology}, 9(1):77--110, 1977.

\bibitem{tufte1983}
E.~Tufte.
\newblock {\em Visual Display of Quantitative Data}.
\newblock 1983.

\bibitem{wang2019datashot}
Y.~Wang, Z.~Sun, H.~Zhang, W.~Cui, K.~Xu, X.~Ma, and D.~Zhang.
\newblock Datashot: Automatic generation of fact sheets from tabular data.
\newblock {\em IEEE Trans. on visualization and computer graphics}, 2019.

\bibitem{Wang2020data}
Z.~{Wang}, J.~{Ritchie}, J.~{Zhou}, F.~{Chevalier}, and B.~{Bach}.
\newblock Data comics for reporting controlled user studies in human-computer
  interaction.
\newblock {\em IEEE Trans. on Visualization and Computer Graphics}, pages 1--1,
  2020.

\bibitem{wang2019comparing}
Z.~Wang, S.~Wang, M.~Farinella, D.~Murray-Rust, N.~Henry~Riche, and B.~Bach.
\newblock Comparing effectiveness and engagement of data comics and
  infographics.
\newblock In {\em Proc. of the ACM Conference on Human Factors in Computing
  Systems}, pages 253:1--253:12, 2019.

\bibitem{wongsuphasawat2015voyager}
K.~Wongsuphasawat, D.~Moritz, A.~Anand, J.~Mackinlay, B.~Howe, and J.~Heer.
\newblock Voyager: Exploratory analysis via faceted browsing of visualization
  recommendations.
\newblock {\em IEEE Trans. on Visualization and Computer Graphics},
  22(1):649--658, 2015.

\bibitem{wongsuphasawat2016towards}
K.~Wongsuphasawat, D.~Moritz, A.~Anand, J.~Mackinlay, B.~Howe, and J.~Heer.
\newblock Towards a general-purpose query language for visualization
  recommendation.
\newblock In {\em Proc. of the ACM Workshop on Human-In-the-Loop Data
  Analytics}, pages 4:1--4:6, 2016.

\bibitem{xia2018dataink}
H.~Xia, N.~H. Riche, F.~Chevalier, B.~D. Araujo, and D.~Wigdor.
\newblock Dataink: Direct and creative data-oriented drawing.
\newblock In {\em Proc. of the ACM Conference on Human Factors in Computing
  Systems}, 2018.

\bibitem{xu2018chart}
S.~Xu, C.~Bryan, J.~K. Li, J.~Zhao, and K.-L. Ma.
\newblock Chart constellations: Effective chart summarization for collaborative
  and multi-user analyses.
\newblock {\em Computer Graphics Forum}, 37(3):75--86, 2018.

\bibitem{Zhao2020chartseer}
J.~{Zhao}, M.~{Fan}, and M.~{Feng}.
\newblock Chartseer: Interactive steering exploratory visual analysis with
  machine intelligence.
\newblock {\em IEEE Trans. on Visualization and Computer Graphics}, pages 1--1,
  2020.

\bibitem{Zhao2017supporting}
J.~Zhao, M.~Glueck, P.~Isenberg, F.~Chevalier, and A.~Khan.
\newblock Supporting handoff in asynchronous collaborative sensemaking using
  knowledge-transfer graphs.
\newblock {\em IEEE Trans. on visualization and computer graphics},
  24(1):340--350, 2017.

\bibitem{zhao2015data}
Z.~Zhao, R.~Marr, and N.~Elmqvist.
\newblock Data comics: Sequential art for data-driven storytelling.
\newblock Technical report, University of Maryland, 2015.

\end{thebibliography}


%



\begin{IEEEbiographynophoto}{Jian Zhao}
is an assistant professor in the Cheriton School of Computer Science at the University of Waterloo, where he directs the WatVis (Waterloo Visualization) group. His research interests include information visualization, human-computer interaction, and data science. His work contributes to the development of advanced interactive visualizations that promote the interplay of human, machine, and data. 
\end{IEEEbiographynophoto}

\begin{IEEEbiographynophoto}{Shenyu Xu}
is a Ph.D. student in computer science at Georgia Institute of Technology, where he works with Prof. Alex Endert. He is a member of the GaTech Visual Analytics Lab. His research lies on the intersection of human-computer interaction and visual analytics. 
\end{IEEEbiographynophoto}

\begin{IEEEbiographynophoto}{Senthil Chandrasegaran}
is an assistant professor in the Faculty of Industrial Design Engineering at the Delft University of Technology. His research focuses on the integration of computer support tools to aid collaboration in early design, and the use of visualization and visual analytics techniques to understand how designers work together. 
\end{IEEEbiographynophoto}

\begin{IEEEbiographynophoto}{Chris Bryan} 
is an assistant professor in the School of Computing, Informatics, and Decision Systems Engineering at the Arizona State University, where he directs the Sonoran Visualization Laboratory (SVL @ ASU). He received the PhD degree in computer science from the University of California, Davis, in 2018. His research areas include information visualization, human-computer interaction, and virtual reality. 
\end{IEEEbiographynophoto}

\begin{IEEEbiographynophoto}{Fan Du}
is a research scientist at Adobe Research. He works on human-centered AI with a focus on empowering data analysts with visualization and machine learning technologies. He develops and transfers research prototypes to Adobe's digital marketing products, including Analytics, Audience Manager, and Experience Platform.  
\end{IEEEbiographynophoto}

\begin{IEEEbiographynophoto}{Aditi Mishra}
is a Ph.D. student in computer science at the Arizona State University, where she works with Prof. Chris Bryan. She is a member of the Sonoran Visualization Laboratory. Her research focuses visual analytics and human-computer interaction.  
\end{IEEEbiographynophoto}

\begin{IEEEbiographynophoto}{Xin Qian}
is a Ph.D. student in the College of Information Studies at the University of Maryland, where she works with Prof. Joel Chan. She is a member of the HCIL lab. Her research focus on natural language processing and human-computer interaction.   
\end{IEEEbiographynophoto}

\begin{IEEEbiographynophoto}{Yiran Li}
is a Ph.D. student in computer science at the University of California, Davis, where she works with Prof. Kwan-Liu Ma. She is a member of the VIDI lab. Her research focuses on visual analytics and information visualization. 
\end{IEEEbiographynophoto}

\begin{IEEEbiographynophoto}{Kwan-Liu Ma} 
is a distinguished professor of computer science at the University of California, Davis. He directs VIDI Labs and UC Davis Center of Excellence for Visualization. Professor Ma received his PhD degree in computer science from the University of Utah in 1993. His research interests include visualization, computer graphics, high-performance computing, and human-computer interaction. Professor Ma was a recipient of the NSF PECASE award in 2000 and the IEEE VGTC 2013 Visualization Technical Achievement Award. He is an IEEE Fellow. He presently serves as the AEIC of IEEE CG\&A.
\end{IEEEbiographynophoto}


\vfill


\end{document}